\documentclass[trackchanges]{aastex701}

\renewcommand{\baselinestretch}{1.2}
\usepackage{tabularx}  
\usepackage{array}     
\usepackage{makecell}
\usepackage{amsmath} 
\usepackage{comment}
\usepackage[table]{xcolor}
\usepackage{graphicx}
\usepackage{float}
\usepackage{placeins}

\shortauthors{A. Gupta et al.}

\begin{document}


\title{Insights on Three Distinct Types of Energetic Particle Flux Variations Ahead of ICME-driven Shocks Observed by ASPEX on board Aditya-L1}


\author[orcid=0009-0000-9917-2694,sname='Gupta']{Aakash Gupta}
\altaffiliation{Physical Research Laboratory (PRL), Ahmedabad - 380009, India}
\affiliation{Physical Research Laboratory (PRL), Ahmedabad - 380009, India}
\affiliation{Department of Physics, Indian Institute of Technology (IIT), Gandhinagar - 382055, India}
\email[show]{aakashgupta.du@gmail.com}  

\author[orcid=0000-0003-2693-5325, sname='Chakrabarty']{Dibyendu Chakrabarty} 
\affiliation{Physical Research Laboratory (PRL), Ahmedabad - 380009, India}
\email{dipu@prl.res.in}

\author[orcid=0000-0001-6018-9018, sname='Yogesh']{Yogesh} 
\affiliation{Department of Physics and Astronomy, University of Iowa, Iowa City IA 52242, USA}
\email{yphy22@gmail.com}

\author[orcid=0000-0001-5742-9033, sname='Kumari']{Anshu Kumari}
\affiliation{Udaipur Solar Observatory (USO), Physical Research Laboratory (PRL), Udaipur – 313001, India}
\email{anshu@prl.res.in}

\author[orcid=0009-0004-7817-0859,sname='Parashar']{Shivam Parashar}
\affiliation{Physical Research Laboratory (PRL), Ahmedabad - 380009, India}
\affiliation{Department of Physics, Indian Institute of Technology (IIT), Gandhinagar - 382055, India}
\email{parasharshivam820@gmail.com} 

\author[orcid=0009-0004-5904-5833, sname='Sebastian']{Jacob Sebastian}
\affiliation{Physical Research Laboratory (PRL), Ahmedabad - 380009, India}
\email{jacobs@prl.res.in}

\author[orcid=0009-0002-9730-3640, sname='Mishra']{Amrit S. Mishra}
\affiliation{Physical Research Laboratory (PRL), Ahmedabad - 380009, India}
\email{amrit@prl.res.in}

\author[orcid=0000-0001-8993-9118, sname='Dalal']{Bijoy Dalal}
\affiliation{Physical Research Laboratory (PRL), Ahmedabad - 380009, India}
\email{bijoydalal.at@gmail.com}

\author[orcid=0000-0002-2050-0913, sname='Vadawale']{Santosh Vadawale}
\affiliation{Physical Research Laboratory (PRL), Ahmedabad - 380009, India}
\email{santoshv@prl.res.in}

\author[orcid=0000-0002-3153-537X, sname='Goyal']{Shiv Kumar Goyal}
\affiliation{Physical Research Laboratory (PRL), Ahmedabad - 380009, India}
\email{goyal@prl.res.in}

\author[orcid=0000-0002-5995-8681, sname='Shanmugam']{M. Shanmugam}
\affiliation{Physical Research Laboratory (PRL), Ahmedabad - 380009, India}
\email{shansm@prl.res.in}

\author[orcid=0000-0003-1693-453X, sname='Bhardwaj']{Anil Bhardwaj}
\affiliation{Physical Research Laboratory (PRL), Ahmedabad - 380009, India}
\email{bhardwaj_spl@yahoo.com}

\author[orcid=0000-0002-1470-8443, sname='Yadav']{Vipin K. Yadav}
\affiliation{Space Physics Laboratory, Vikram Sarabhai Space Centre, Indian Space Research Organization (ISRO), Trivandrum - 695022, India}
\email{vipin_ky@vssc.gov.in}

\author{Srikar P. Tadepalli}
\affiliation{UR Rao Satellite Centre (URSC), Indian Space Research Organization (ISRO), Bengaluru - 560017, India}
\email{srikar@ursc.gov.in}

\begin{abstract}
We investigate upstream energetic-ion transport associated with ten interplanetary coronal mass ejection (ICME)-driven shocks observed near 1 au using multidirectional measurements from the Supra-Thermal and Energetic Particle Spectrometer (STEPS) onboard Aditya-L1. Based on energy-dependent intensity profiles and quantitatively determined onset times, the events are classified into three categories: dispersed, dispersionless, and mixed. We combine pitch-angle anisotropy, minimum variance analysis, reduced magnetic helicity, two-dimensional Grad--Shafranov reconstruction, quasi-linear transport diagnostics, shock parameters, and energetic-ion spectral fits to characterize the physical conditions associated with each category. The seven dispersed events exhibit systematic energy-dependent arrivals, generally intermittent magnetic-helicity signatures, and larger parallel mean free paths than the dispersionless events, consistent with less restricted parallel propagation. In contrast, the two dispersionless events show nearly simultaneous enhancements across multiple energy channels, extended intervals of enhanced reduced magnetic helicity, and well-defined closed magnetic-flux contours in Grad--Shafranov reconstructions. They also exhibit shorter parallel mean free paths, higher pitch-angle diffusion coefficients, and comparatively strong shocks. The mixed event transitions from dispersed arrivals to nearly simultaneous enhancements as the shock approaches, accompanied by evolving anisotropy and distinct scattering properties. Spectral indices range from approximately \(-2.38\) to \(-0.86\) for the dispersed events and from \(-0.82\) to \(-0.26\) for the dispersionless and mixed events, indicating substantially harder spectra in the latter groups. These results reveal systematic associations between energetic-ion transport, magnetic structure, shock conditions, and spectral hardness. They suggest that the observed signatures reflect the combined influence of shock-associated energization and energy-dependent transport, with magnetic topology potentially regulating particle propagation and confinement. 
\end{abstract}


\section{Introduction} \label{sec: Introduction}

Solar energetic particle (SEP) events are among the most important manifestations of space weather in the heliosphere \citep{Reames1999, Gopalswamy2008, Vainio2009}. These events are characterized by sudden enhancements in energetic charged particle intensities, spanning energies from a few tens of keV n$^{-1}$ to several hundreds of MeV n$^{-1}$ , with durations ranging from a few hours to several days. During major SEP events, particle intensities can exceed quiet-time background levels by many orders of magnitude. SEPs are commonly associated with transient solar eruptive phenomena such as solar flares and coronal mass ejections (CMEs), both of which provide favorable environments for efficient particle acceleration through magnetic reconnection and large-scale shock formation in the solar corona and interplanetary medium \citep{Reames1999, Cliver2000}. In situ observations throughout the heliosphere routinely detect energetic ions and electrons over broad energy ranges extending from a few keV to hundreds of MeV and, in extreme events, into the relativistic regime \citep{Desai2016}.

Multiple physical processes have been proposed to explain SEP acceleration. These include particle energization during magnetic reconnection associated with solar flares and jets \citep{Kahler2007, Krucker2011}, diffusive shock acceleration, shock drift acceleration at CME-driven shocks propagating through the corona and interplanetary space \citep{Jokipii1982, Simnett2002, Desai2016, Kouloumvakos2019, Jebaraj2024}, and acceleration occurring within reconnecting current sheets and evolving post eruptive magnetic structures associated with CMEs \citep{Kahler1992, Maia2004, Klein2005}. Recent observations from the Aditya Solar Wind Particle Experiment (ASPEX) onboard Aditya-L1 have further shown that interactions between multiple interplanetary CMEs can generate extended interaction regions associated with shock propagation, particle energization, and redistribution of energetic ions in the heliosphere \citep{Parashar2026}. SEP events are traditionally divided into impulsive and gradual classes based on their temporal evolution, elemental composition, charge states, energy spectra, and associated radio signatures \citep{Cane1986, Reames1999}. Gradual SEP events are often observed over extremely broad heliolongitudinal extents, implying efficient acceleration and large-scale particle transport throughout the heliosphere \citep{Reames1996, Lario2006, Wibberenz2006, Dresing2012, Richardson2014, Papaioannou2014, Paassilta2018, Xie2019, RodriguezGarcia2021, Kouloumvakos2022, Dresing2023, Khoo2024}.

The transport of energetic particles in the heliosphere is strongly governed by the turbulent interplanetary magnetic field (IMF). During propagation, charged particles undergo both parallel and perpendicular diffusion relative to the ambient magnetic field \citep{Jokipii1966b, Giacalone1999, Matthaeus2003, Qin2004, Shalchi2010, Ruffolo2012}. Consequently, SEP intensity profiles measured near Earth frequently exhibit substantial spatial and temporal variability arising from the combined effects of acceleration, scattering, magnetic connectivity, and large-scale heliospheric structures \citep{Guo2010, Giacalone2012, Giacalone2017, Kong2017, Chen2022, Wijsen2023}. The anisotropy of energetic particles serves as an important diagnostic of these transport processes. Strong anisotropies generally indicate weak pitch-angle scattering and efficient field-aligned streaming, whereas near-isotropic distributions imply enhanced scattering and cross-field diffusion \citep{Dresing2014}.


One of the most intriguing transport-related SEP phenomena is the occurrence of transient intensity modulations such as dropouts and dispersion-less particle variations \citep{Mazur2000, Gosling2004}. These events are characterized by abrupt changes in particle intensities with remarkably sharp temporal boundaries occurring simultaneously across broad energy ranges. Similar dispersion-less intensity modulations have previously been reported for both energetic ions and electrons during impulsive SEP events \citep{Anderson1986, Buttighoffer1998, Reames1992, Chollet2008}. Such observations suggest that localized magnetic connectivity and coherent magnetic structures play an important role during particle propagation.

Previous observational and theoretical studies have demonstrated that coherent magnetic structures embedded within solar wind turbulence, including magnetic flux ropes, magnetic islands, current sheets, and magnetic conduit-like structures, can strongly influence energetic particle transport in the heliosphere \citep{Bruno2001, Borovsky2008, Cartwright2010, Trenchi2013a, Trenchi2013b}. These mesoscale magnetic structures can enhance magnetic connectivity, suppress cross-field diffusion, and provide efficient field-aligned transport pathways for energetic particles. Consequently, energetic particle dropouts and dispersion-less flux variations are frequently associated with abrupt changes in magnetic topology and crossings of coherent magnetic structures \citep{Chuychai2005, Trenchi2013a, Trenchi2013b}. In such scenarios, particles can remain confined within localized magnetic flux tubes or magnetic conduits, producing sharp intensity boundaries and nearly simultaneous particle enhancements across broad energy ranges \citep{Trenchi2013a, Trenchi2013b}. 

Magnetic flux rope structures are also frequently observed within CMEs propagating through the heliosphere. Both remote-sensing and in situ observations indicate that many CMEs contain twisted or helical magnetic field configurations enclosing confined plasma \citep{Vourlidas2014}. Such coherent magnetic structures can substantially influence SEP transport by modifying magnetic connectivity, enabling magnetic trapping, and supporting bidirectional streaming along closed or partially closed magnetic field lines \citep{Malandraki2002, Tan2012, Li2020, Wei2024, RodriguezGarcia2025}. Recent observations from Parker Solar Probe and Solar Orbiter further suggest that small scale flux ropes and magnetic conduit like structures may play a critical role in producing localized SEP transport signatures, including dispersion-less intensity variations and bidirectional particle streaming \citep{Dresing2016, GomezHerrero2017, Wimmer2023, Ding2025}.


Despite significant progress in understanding solar energetic particle (SEP) acceleration and transport, the relationship between local heliospheric magnetic field topology and the transport characteristics of energetic particles upstream of interplanetary coronal mass ejection (ICME)-driven shocks remains incompletely understood. Previous studies have largely focused on dispersion-less particle dropouts, particle anisotropy, and magnetic structures independently \citep{Gosling2004,Chollet2008, Blanco2013, Ho2022}. In particular, the extent to which magnetic structures, characterized by Minimum Variance Analysis (MVA) eigenvalue ratios ($\lambda_2/\lambda_3$), Grad–Shafranov (GS) reconstruction of two-dimensional (2D) magnetic flux surfaces, and reduced magnetic helicity ($\sigma_m$), influence the observed energy-time profiles of upstream energetic ions has not been systematically established. Moreover, the relationship between these magnetic properties, particle transport coefficients, and energy spectral evolution remains insufficiently constrained. A comparative investigation of energy-dispersed, dispersion-less, and mixed particle enhancements, using multi-directional measurements across a broad energy range ($\sim 0.12\text{--}1.33\ \mathrm{MeV/n}$), can provide important insights into the physical processes governing energetic ion propagation, scattering, and confinement in complex interplanetary magnetic environments. Resolving these coupled topological, transport, and energization processes is essential for establishing a predictive, physics-based model of SEP transport in complex interplanetary environments.

In this study, we investigate upstream energetic ion enhancements associated with ten ICME-driven shocks observed near the Sun–Earth L1 point using multi-directional measurements from the Supra-Thermal and Energetic Particle Spectrometer (STEPS) onboard the Aditya Solar Wind Particle Experiment (ASPEX) on Aditya-L1. Based on their energy-time signatures, the observed enhancements are classified into three transport regimes: dispersed, dispersion-less, and mixed. We characterize the particle distributions through pitch-angle anisotropy and energy spectral index analyses and investigate their association with local magnetic field geometry and topology using MVA eigenvalue ratios, GS-based 2D flux-surface reconstruction, and reduced magnetic helicity. We further estimate particle transport coefficients to examine how the inferred transport properties relate to magnetic field organization and turbulence. By combining these complementary diagnostics, we assess the extent to which magnetic structure and particle transport characteristics distinguish the three observational regimes and explore their implications for energetic ion propagation and possible energization near ICME-driven shocks. The paper is organized as follows. Section 2 describes the instrumentation and data sets used in this study. Section 3 presents the observations and detailed methodology, including the event classification based on visual inspection and quantitative velocity dispersion analysis, estimation of energetic-ion spectral indices, pitch-angle distributions and first-order anisotropy, magnetic topology analysis using reduced magnetic helicity, two-dimensional Grad–Shafranov reconstruction and minimum variance analysis (MVA), estimation of ICME shock parameters, and calculation of energetic-ion transport coefficients. Section 4 presents and discusses the results for the dispersionless, dispersed, and mixed upstream energetic-ion events. Section 5 summarizes the principal findings and discusses their implications for energetic-particle transport and shock-associated energization in the heliosphere. The energy spectra for the dispersed events and for the dispersionless and mixed events are provided in the appendices, and the data availability statement is included at the end of the paper.

\section{Instrumentation and Dataset} \label{sec:Instrumentation and Dataset}

In this study, we utilize data from the Solar Wind Ion Spectrometer (SWIS) \citep{PKumar2025, Akumar2026} and Supra-Thermal and Energetic Particle Spectrometer (STEPS) \citep{Goyal_et_al_2025, Sebastian2026}, both are subsystem of the Aditya Solar wind Particle Experiment (ASPEX) \citep{Janardhan_et_al_2017, Goyal2018, PKumar2025, Jsebastian_et_al_2025} onboard Aditya-L1 \citep{Seetha2017, Tripathi2022, Parate2025}. Aditya-L1 is India’s first observatory class mission dedicated to the study of the Sun and heliosphere. The mission was launched on 2 September 2023 and was inserted into a halo orbit around the first Lagrange point (L1) of the Sun–Earth system on 6 January 2024. SWIS measures solar wind ions in the energy range spanning form 100 eV - 20 KeV using two electrostatic analyzers, top hat analyzer 1 and top hat analyzer 2 (THA-1 and THA-2) and provides the bulk parameters of the solar wind such as density, temperature and bulk velocity.  STEPS is designed to investigate energetic ion populations in the heliosphere. The instrument consists of six detector units, namely Sun-Radial (SR), Inter-Mediate (IM), Parker Spiral (PS), North Pointing (NP), South Pointing (SP), and Earth Pointing (EP). Due to persistent light saturation effects in the SR and SP units, observations from these detectors are not included in the present analysis. As discussed in \citet{Goyal_et_al_2025, Gupta2025}, and \citet{Sebastian2026}, the NP and IM detectors employ a single-window configuration with a dead layer thickness of $0.2\,\mu\mathrm{m}$. In comparison, the PS and EP units utilize a composite detector arrangement comprising a dual-window Si-PIN (Silicon Positive-Intrinsic-Negative) detector together with a scintillator and silicon photomultiplier (SiPM) assembly. The dual-window Si-PIN detector contains two distinct sensing regions with different dead layer thicknesses. The inner circular region (PS-Inn and EP-Inn), having a diameter of $7\, \mathrm{mm}$, is characterized by a dead layer thickness of $0.1\,\mu\mathrm{m}$, whereas the surrounding annular region (PS-Out and EP-Out), covering diameters from $7\,\mathrm{mm}$ to $18\,\mathrm{mm}$, has a thicker dead layer of $0.8\,\mu\mathrm{m}$. The present work makes use of species-integrated energetic ion flux measurements obtained specifically from the PS-Out, EP-Out, IM and NP detector units. Therefore, any mention of the PS and EP units throughout this study refers solely to the PS-Out and EP-Out detectors, respectively. The detectors of ASPEX-STEPS incorporate dedicated electron-rejection capability, which effectively suppresses contamination from energetic electrons up to approximately 1 MeV, thereby ensuring that the measured counts predominantly correspond to ions \citep{Goyal_et_al_2025, Sebastian2026}. Among the four detector units, the PS, IM, and EP detectors primarily measure ions within the X-Y plane of the Geocentric Solar Ecliptic (GSE) coordinate system, while the NP detector is oriented approximately along the positive Z-axis. STEPS measures the flux of energetic ions in the energy range extending from 20 keV/n to 6 MeV/n. We use the linear (in logarithmic flux–energy space) part of the spectra to derive the spectral indices from the fluxes. This leads to selection of ion fluxes with lower cut-off of 0.3 MeV/n for PS and EP detectors and 0.12 MeV/n for IM and NP detectors. In addition to the ASPEX-STEPS observations, this study also incorporates interplanetary magnetic field (IMF) measurements from the Magnetometer (MAG) payload \citep{Yadav2025} on-board Aditya-L1, available at a temporal resolution of 10 seconds. In addition to the ASPEX and MAG instruments onboard Aditya-L1, we utilized solar wind bulk parameters measured by the Solar Wind Experiment (SWE; \citet{Ogilvie1995}) and magnetic field measurements from the Magnetic Field Investigation (MFI; \citet{Lepping1995}), both onboard NASA's Wind spacecraft \citep{Acuna1995}.


\section{Observations and Methodology} \label{sec:Observations and Methodology}
In this study, we investigate ten ICME events selected from the Wind ICME catalog\footnote{\url{https://wind.nasa.gov/ICME_catalog/ICME_catalog_viewer.php}}
together with the NASA DONKI database\footnote{\url{https://kauai.ccmc.gsfc.nasa.gov/DONKI/search/}}. More ICME-driven shock events occurred during the ASPEX-STEPS operational period; however, only events with (i) a clear shock signature in the ASPEX-STEPS observations and (ii) simultaneous availability of magnetic field measurements from the Aditya-L1 magnetometer (MAG) were included in this study. A comprehensive description of the Wind ICME catalog, including the criteria adopted for ICME identification, is discussed in \cite{Chinchilla2019}. For each selected ICME event, temporal variations of energetic ion fluxes are examined using the PS, IM, EP, and NP detectors of the ASPEX-STEPS instrument onboard Aditya-L1. The corresponding energy ranges considered for the analysis are $0.36$--$1.32$ MeV/n, $0.14$--$1.22$ MeV/n, $0.39$--$1.33$ MeV/n, and $0.12$--$1.23$ MeV/n for the PS, IM, EP, and NP detectors, respectively. During ICME passage at L1, enhancements in energetic ion flux are frequently observed ahead of the associated shock arrival. Understanding the origin and evolution of these upstream particle enhancements requires information on both the energetic particle distributions and the surrounding magnetic-field configuration. In this context, the halo orbit of Aditya-L1 around the Sun-Earth L1 point, together with the presence of other heliospheric observatories such as IMAP, Wind, ACE, DSCOVR, and SOLAR-1, provides opportunities for coordinated multi-spacecraft investigations. Combining the multi-directional energetic ion measurements from ASPEX-STEPS with complementary in situ observations can help distinguish spatial variations in magnetic structures from temporal variations in shock-associated particle acceleration, thereby providing additional constraints on the global magnetic connectivity of the shock \citep[e.g.,][]{Zhao2027}. Motivated by this, we focus on the energetic ion enhancements observed upstream of each shock. Specifically, the interval extending from the onset of the energetic ion flux enhancement to the shock arrival time, typically corresponding to the interval in which the ion flux reaches its maximum, is identified as the upstream energetic ion enhancement region. This region is indicated by the gray shading in Figure~1. The observed enhancements are investigated as signatures of particle energization and transport in the upstream region ahead of the shock front.

\subsection{Event Classification Based on Visual Inspection and Quantitative Velocity Dispersion Analysis} \label{sec:event_classification}
The temporal evolution of these upstream energetic ion enhancements reveals distinct energy-dependent behaviors, which form the basis for the classification of the events into three different categories. (1) In seven out of the ten ICME events, the upstream energetic ion flux enhancements occur in an energy-dispersed manner, where higher-energy particles are observed earlier than lower-energy particles. These events are classified as \textit{dispersed events}. (2) In two ICME events, the flux enhancement is observed nearly simultaneously across all energy channels without any clear energy dispersion. Such events are categorized as \textit{dispersion-less events}. (3) In one ICME event, the upstream energetic ion enhancement initially exhibits a dispersed behavior and subsequently transitions into a dispersion-less enhancement near the shock arrival time. This event is classified as a \textit{mixed event}, representing characteristics of both dispersed and dispersion-less flux variations. Figure 1 presents representative examples of these three distinct types of temporal variations observed in energetic ions measurements from ASPEX-STEPS onboard Aditya-L1. 

\begin{figure*}[ht!]
\centering
\includegraphics[width=\textwidth]{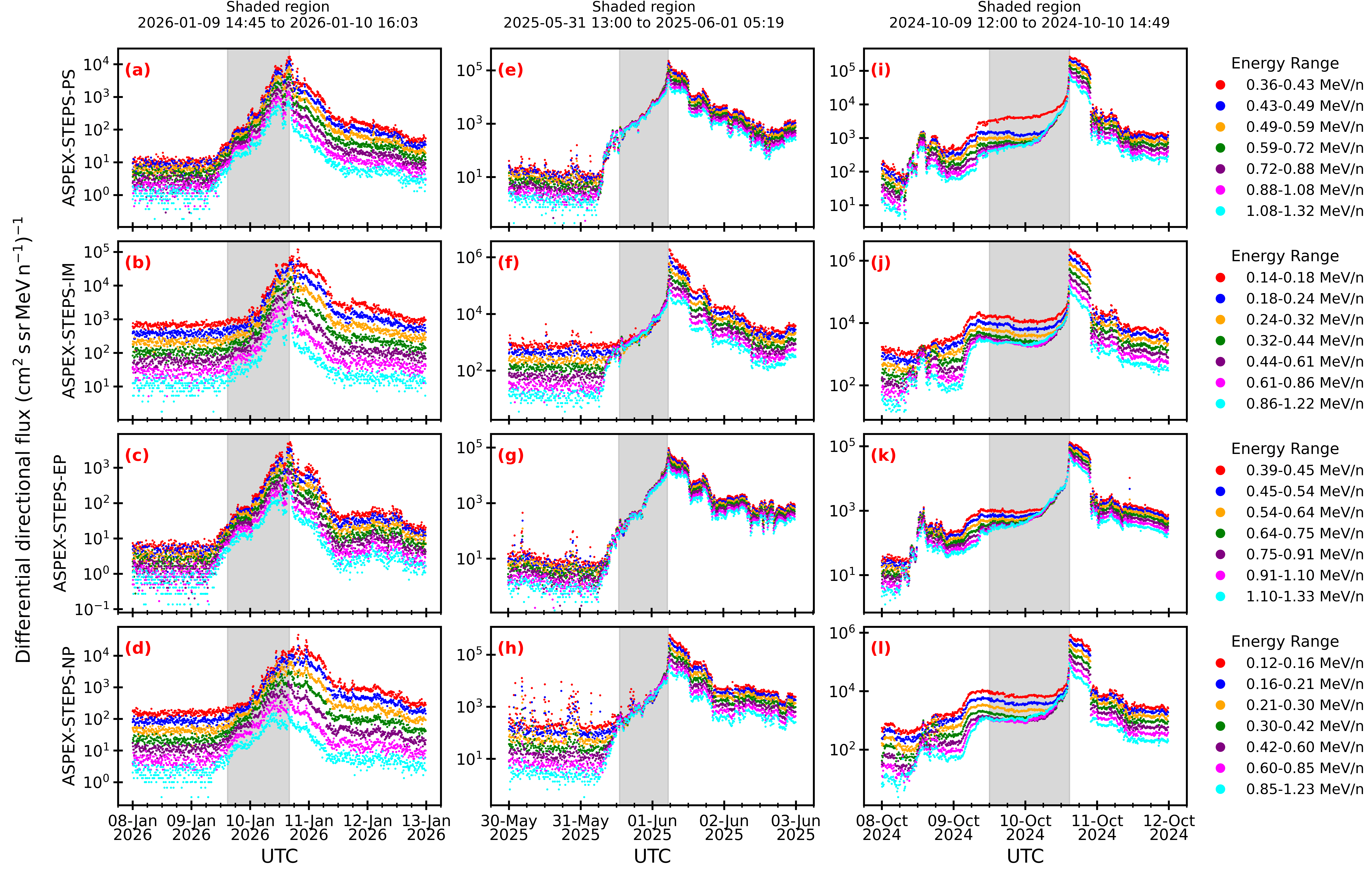}
\caption{This figure represents the example of three distinct types of upstream energetic ions flux variations observed ahead of ICME-driven shocks at L1: dispersed (left column), dispersion-less (middle column), and mixed (right column) events. Panels (a), (e), and (i) show energetic ion fluxes measured by the Parker Spiral (PS) detector of ASPEX-STEPS; panels (b), (f), and (j) correspond to the Intermediate (IM) detector; panels (c), (g), and (k) show measurements from the Earth Pointing (EP) detector; and panels (d), (h), and (l) present observations from the North Pointing (NP) detector. The colored symbols represent different ion energy channels, with the corresponding energy ranges listed to the right of each row. The grey shaded regions indicate the upstream energetic ion enhancement intervals. The dispersed event exhibits clear energy-dependent temporal dispersion, where higher-energy ions arrive earlier than lower-energy ions. In contrast, the dispersion-less event shows nearly simultaneous flux enhancements across all energy channels. The mixed event initially displays dispersed behavior followed by a transition to dispersion-less enhancement close to the shock arrival.
}
\label{fig:general}
\end{figure*}

\begin{figure*}[ht!]
\centering
\includegraphics[width=\textwidth]{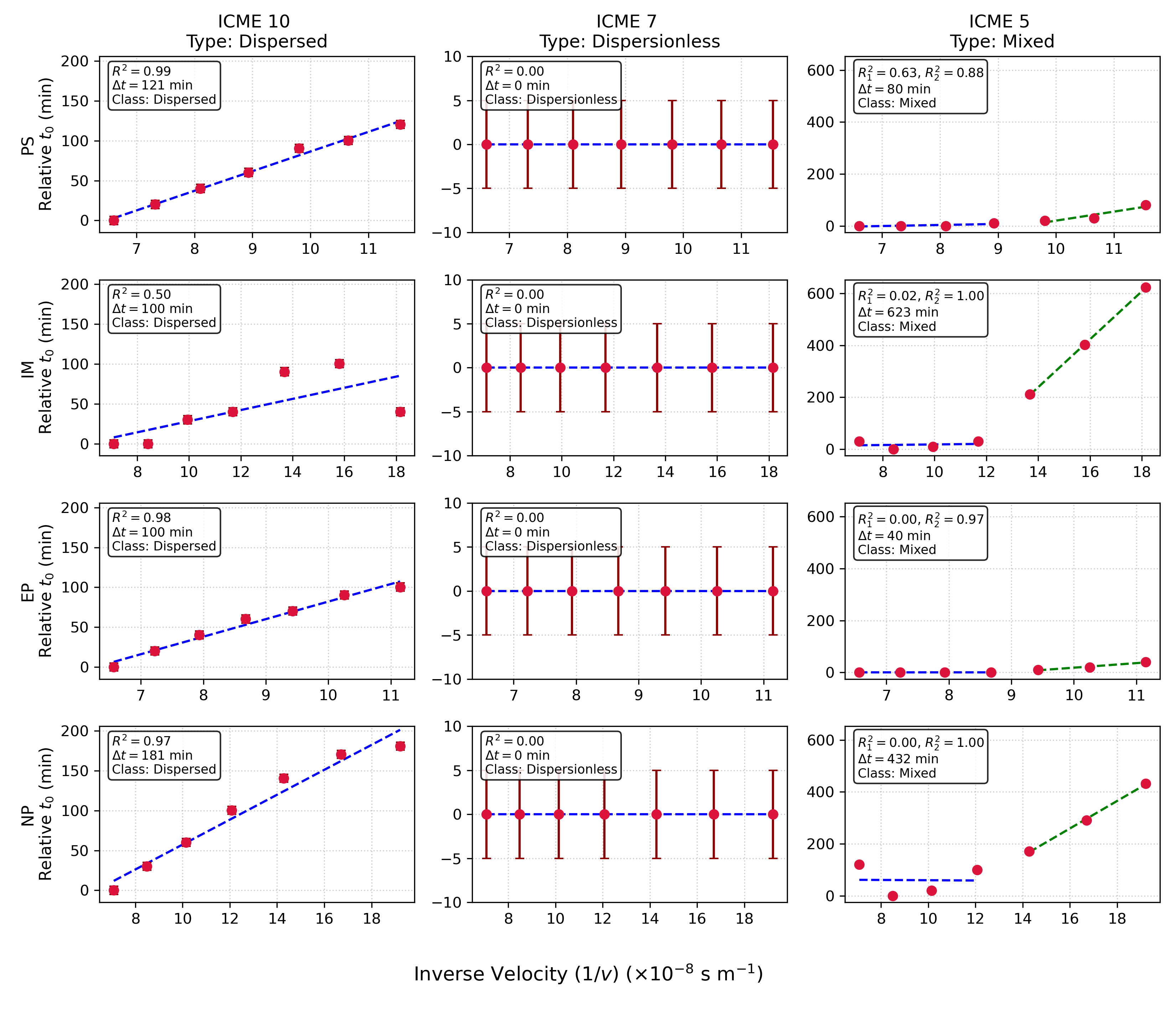}
\caption{Velocity dispersion analysis (VDA) results showing relative onset time ($t_0$) versus inverse particle velocity ($1/v$) for three representative ICME events across the four ASPEX-STEPS detector directions (PS, IM, EP, and NP from top to bottom): ICME 10 (Dispersed class; left column), ICME 7 (Dispersionless class; middle column), and ICME 5 (Mixed class; right column). Red circles denote the derived energy-dependent onset times with an uncertainty of $\pm 5$ min. Dashed lines represent linear regression fits used to characterize the arrival-time behavior, with the corresponding coefficient of determination ($R^2$), total onset-time spread ($\Delta t$), and event classification displayed in each panel legend.}
\label{fig:general}
\end{figure*}

To quantitatively validate the event classifications, we performed a velocity dispersion analysis (VDA) using the energy-dependent onset times derived separately for the PS, IM, EP, and NP detector measurements of ASPEX-STEPS. For each energy channel, the pre-event background was estimated as the mean flux over a selected quiet interval and subtracted from the flux measured during the upstream enhancement interval. Negative background-subtracted flux values were set to zero, and the resulting net flux profiles were smoothed using a Savitzky--Golay filter \citep{SavitzkyGolay1964}. The onset time, $t_0(E_i)$, was defined as the time at which the cumulative sum of the smoothed net flux reached 3\% of the total net fluence over the selected interval. An onset-time uncertainty of $\pm 5$ min was adopted to account for the temporal resolution of the measurements from ASPEX-STEPS. For each detector direction, the total observed onset-time spread was calculated as $\Delta t = t_{0,\max} - t_{0,\min}$, and the onset times were examined as a function of inverse particle velocity ($1/v$), with particle velocities calculated relativistically from the mean energies of the corresponding channels.  Linear regression of $t_0$ against $1/v$ was used to characterize the energy-dependent arrival-time behavior, following the general VDA approach used in previous solar energetic particle studies \citep{HuttunenHeikinmaa2005,Palmroos2025}. The event classes were distinguished using the onset-time distributions and their temporal clustering. An enhancement was classified as \textit{dispersion-less} when the observed onset-time spread, $\Delta t=t_{0,\max}-t_{0,\min}$, was $\leq 15$ min and the onsets showed substantial clustering within a 10-min window, corresponding to at least 60\% of the available channels or at least four channels, as applicable. An enhancement was classified as \textit{dispersed} when the onset-time spread was $\geq 20$ min and the onset times followed a clear energy-dependent sequence, with no more than two channel onsets occurring within the same 10-min window. Enhancements exhibiting intermediate or combined behavior, including a cluster of earlier higher-energy onsets followed by delayed lower-energy onsets or multiple onset clusters extending over a longer interval, were classified as \textit{mixed}. These criteria provide a quantitative basis for distinguishing the three classes rather than relying solely on visual inspection of the energy--time profiles. The results for the three representative events shown in Figure~1 are presented in Figure~2, illustrating the onset-time distributions and their differences across the four detector directions.

\subsection{Estimation of Spectral Indices for all ICME events}
\label{sec:spectral_indices}
 The spectral indices for these upstream energetic ions, for all ten ICME events, are determined by fitting the ion spectra with a power-law function of the form \( J = A E^{-m} \). The fitting procedure is performed in logarithmic flux–energy space by considering only the linear portion of the spectra within the selected energy intervals for the PS, IM, EP, and NP detectors of ASPEX-STEPS (for more details see \cite{Gupta2025}). The fitted spectra and the corresponding spectral index values for the dispersed events are presented in Appendix Figure A1, whereas those for the dispersion-less and mixed events are shown in Appendix Figure A2. The start and end times of the upstream energetic ion enhancement intervals preceding the selected ICME-driven shocks at the L1 point, together with the corresponding shock arrival times and the derived spectral index values for the PS, IM, EP, and NP detectors, are summarized in Table 1. The end time of the upstream enhancement interval corresponds to the shock arrival time at L1. Table 1 shows that the seven dispersed events exhibit relatively soft power-law spectra, with spectral indices ranging from $-0.86 \pm 0.07$ to $-2.38 \pm 0.07$ across the four ASPEX-STEPS viewing directions. In contrast, the two dispersion-less events (Events 2 and 7; in table 1) and the mixed event (Event 5; in table 1) display significantly harder spectra, with spectral indices ranging from $-0.23 \pm 0.06$ to $-0.82 \pm 0.16$. These considerably flatter spectral slopes indicate an enhanced contribution of high-energy particles relative to the dispersed events. Notably, the dispersion-less and mixed events exhibit spectral indices substantially harder than those observed for the dispersed events. Several of these values have magnitudes smaller than unity, which are uncommon for upstream energetic ions spectra associated with ICME-driven shocks. These unusually hard spectra indicate that the underlying particle acceleration and transport processes differ from those operating during the dispersed events and may involve additional mechanisms beyond classical diffusive transport. Their physical implications are discussed in Section 5.

\begin{deluxetable*}{cccccccc}
\tabletypesize{\scriptsize}
\tablecolumns{8}
\tablecaption{Start and end times of the upstream energetic ion enhancement intervals associated with the selected ICME-driven shocks, the corresponding shock arrival times at the L1 point, flux variation types, and spectral indices derived from the PS, IM, EP, and NP detectors of ASPEX-STEPS onboard Aditya-L1. The ten events are classified into three distinct types: dispersion-less events (Events 2 and 7), dispersed events (Events 1, 3, 4, 6, 8, 9, and 10), and a mixed event (Event 5). \label{tab:icme_si}}

\tablehead{
\colhead{ICME} & \colhead{Start date} & \colhead{End date} & \colhead{Type of} &
\multicolumn{4}{c}{Spectral Index for ASPEX-STEPS} \\
\colhead{No.} & \colhead{(YYYY-MM-DD HH:MM)} & \colhead{(YYYY-MM-DD HH:MM)} & \colhead{flux variation} &
\colhead{PS} & \colhead{IM} & \colhead{EP} & \colhead{NP}
}

\startdata
1  & 2024-02-06 13:23 & 2024-02-11 01:28 & Dispersed       & $-1.72 \pm 0.07$ & $-1.58 \pm 0.12$ & $-1.64 \pm 0.08$ & $-1.56 \pm 0.12$ \\
2  & 2024-03-24 00:00 & 2024-03-24 14:10 & Dispersion-less & $-0.79 \pm 0.02$ & $-0.82 \pm 0.02$ & $-0.64 \pm 0.01$ & $-0.76 \pm 0.01$ \\
3  & 2024-07-22 16:07 & 2024-07-23 20:00 & Dispersed       & $-1.90 \pm 0.01$ & $-1.54 \pm 0.09$ & $-2.38 \pm 0.07$ & $-1.69 \pm 0.01$ \\
4  & 2024-09-17 03:41 & 2024-09-17 10:08 & Dispersed       & $-1.53 \pm 0.01$ & $-1.66 \pm 0.03$ & $-1.38 \pm 0.01$ & $-1.55 \pm 0.01$ \\
5  & 2024-10-09 12:00 & 2024-10-10 14:49 & Mixed           & $-0.71 \pm 0.16$ & $-0.81 \pm 0.09$ & $-0.26 \pm 0.03$ & $-0.55 \pm 0.10$ \\
6  & 2025-02-25 08:28 & 2025-02-26 02:47 & Dispersed       & $-1.29 \pm 0.02$ & $-1.15 \pm 0.03$ & $-1.22 \pm 0.02$ & $-1.16 \pm 0.03$ \\
7  & 2025-05-31 13:00 & 2025-06-01 05:19 & Dispersion-less & $-0.58 \pm 0.03$ & $-0.23 \pm 0.06$ & $-0.49 \pm 0.04$ & $-0.28 \pm 0.07$ \\
8  & 2025-08-31 18:34 & 2025-09-01 20:28 & Dispersed       & $-1.50 \pm 0.04$ & $-1.18 \pm 0.07$ & $-1.17 \pm 0.04$ & $-0.97 \pm 0.07$ \\
9  & 2025-11-11 06:00 & 2025-11-12 00:00 & Dispersed       & $-1.20 \pm 0.04$ & $-0.94 \pm 0.05$ & $-1.08 \pm 0.04$ & $-0.86 \pm 0.07$ \\
10 & 2026-01-09 14:45 & 2026-01-10 16:03 & Dispersed       & $-1.53 \pm 0.05$ & $-1.12 \pm 0.07$ & $-1.57 \pm 0.06$ & $-1.17 \pm 0.08$ \\
\enddata
\end{deluxetable*}


\subsection{Pitch-Angle Distributions and First-Order Anisotropy of Energetic Ions}
\label{sec:pitch_angle_anisotropy}
To further investigate the dispersed, dispersionless, and mixed events, the pitch angles of energetic ions were calculated for the PS, IM, EP, and NP telescopes of ASPEX-STEPS using magnetic field measurements from the Aditya-L1 Magnetometer (MAG). The pitch angle, $\alpha_k$, is defined as the angle between the particle velocity direction associated with the $k$-th telescope and the local interplanetary magnetic field direction:

\begin{equation}
\alpha_k = \cos^{-1} \left(
\frac{\mathbf{L}_k \cdot \mathbf{B}}
{|\mathbf{L}_k|\,|\mathbf{B}|}
\right),
\label{eq:pitch_angle}
\end{equation}

where $\mathbf{B}$ is the magnetic field vector and $\mathbf{L}_k$ is the unit vector representing the assumed particle-flow direction associated with the $k$-th telescope. Since $|\mathbf{L}_k|=1$, the pitch-angle cosine is given by $\mu_k=\cos\alpha_k=(\mathbf{L}_k\cdot\mathbf{B})/|\mathbf{B}|$. The telescope viewing directions were specified using their nominal angles in the RTN plane.

In addition, the first-order anisotropy index ($A$) \citep{Jokipii1966b, Sanderson1985} was calculated to quantify the net streaming of energetic ions along the interplanetary magnetic field (IMF) during the dispersed, dispersionless, and mixed events. The calculation combines the energy-integrated ion fluxes measured by the four ASPEX-STEPS telescopes (PS, IM, EP, and NP) with magnetic field measurements from the Aditya-L1 Magnetometer (MAG). The magnetic field measurements were resampled into 10-minute intervals, and the corresponding standard errors of the mean were used to estimate the uncertainties in the magnetic field components. The STEPS fluxes also have cadence of 10 minutes for the same intervals and integrated over the considered energy channels using their respective energy-bin widths to obtain the energy-integrated flux, $J_k$, for each telescope. Because the four STEPS telescopes provide measurements along four discrete viewing directions rather than continuous coverage of the full pitch-angle range, the first-order anisotropy was estimated using a discrete approximation to the pitch-angle integral. For each telescope, the pitch-angle cosine, $\mu_k=\cos\alpha_k$, was calculated using equation 1. The anisotropy index was then calculated as

\begin{equation}
A = 3 \frac{\displaystyle\sum_{k=1}^{4} J_k\mu_k}
{\displaystyle\sum_{k=1}^{4} J_k},
\label{eq:anisotropy}
\end{equation}

where $J_k$ and $\mu_k$ denote the energy-integrated flux and pitch-angle cosine for the $k$-th telescope, respectively. Since the telescope viewing directions do not necessarily provide uniform pitch-angle coverage, the resulting anisotropy should be interpreted as an estimate based on the available directional sampling rather than an exact integration over a continuously measured pitch-angle distribution.

The uncertainty in $A$ was estimated by propagating the uncertainties in the energy-integrated fluxes and pitch-angle cosines. The flux uncertainty, $\sigma_{J_k}$, was calculated by propagating the standard errors of the mean fluxes across the energy channels, weighted by their respective energy-bin widths. The uncertainty in $\mu_k$, denoted by $\sigma_{\mu_k}$, was obtained by propagating the standard errors of the magnetic field components through the expression for the pitch-angle cosine. Assuming independent uncertainties among the four telescope fluxes and pitch-angle estimates, the resulting uncertainty in the anisotropy index is

\begin{equation}
\begin{aligned}
\sigma_A^2 ={}&
\sum_{k=1}^{4}
\left(\frac{3\mu_k-A}{D}\right)^2
\sigma_{J_k}^{2} +
\sum_{k=1}^{4}
\left(\frac{3J_k}{D}\right)^2
\sigma_{\mu_k}^{2},
\end{aligned}
\label{eq:anisotropy_uncertainty}
\end{equation}

where $D=\sum_{k=1}^{4}J_k$. The two terms account for the contributions from flux and pitch-angle uncertainties, respectively. Positive and negative values of $A$ indicate preferential streaming along and opposite to the local magnetic field direction, respectively, whereas values close to zero indicate weak net first-order streaming. Temporal variations and sign reversals in $A$ were examined to investigate changes in the directional distribution of energetic ions.


\subsection{Magnetic Topology: Reduced Magnetic Helicity, Grad--Shafranov Reconstruction, and MVA Hodograms}
\label{sec:magnetic_topology}

To investigate the magnetic topology associated with energetic-ion transport during ICME passages, we employed two complementary approaches: reduced magnetic helicity analysis and Grad--Shafranov (GS) reconstruction. Reduced magnetic helicity characterizes the scale-dependent polarization and rotational properties of magnetic-field fluctuations, whereas GS reconstruction provides a two-dimensional representation of the magnetic-field structure under the assumptions of approximate time stationarity and translational symmetry.

\subsubsection{Reduced Magnetic Helicity}
\label{sec:reduced_helicity}

The reduced magnetic helicity, $\sigma_m$, was calculated using the Morlet wavelet transform of the magnetic-field components in the radial--tangential--normal (RTN) coordinate system, following the methods described by \citealt{Zhao2020,Zhao2021a,Pal2025}. It is defined as

\begin{equation}
\sigma_m(s,t) =
\frac{2\,\mathrm{Im}\left[
\tilde{B}_T^{*}(s,t)\tilde{B}_N(s,t)
\right]}
{|\tilde{B}_R(s,t)|^2+
|\tilde{B}_T(s,t)|^2+
|\tilde{B}_N(s,t)|^2},
\label{eq:reduced_magnetic_helicity}
\end{equation}

where $\tilde{B}_R$, $\tilde{B}_T$, and $\tilde{B}_N$ are the complex Morlet wavelet coefficients of the RTN magnetic-field components, the asterisk denotes the complex conjugate, $\mathrm{Im}$ represents the imaginary part, and $s$ is the wavelet scale. The reduced magnetic helicity quantifies the degree of magnetic-field polarization and rotation at a given scale and time. Values of $|\sigma_m|$ approaching unity indicate strongly polarized fluctuations, while values near zero indicate weak net polarization in the measured components. Following previous studies \citep{Matthaeus1982,Zhao2021a,Zhao2021b}, we use $|\sigma_m|>0.6$ as an empirical indicator of coherent, flux-rope-like magnetic structure. However, reduced magnetic helicity is a single-spacecraft diagnostic and does not, by itself, provide a complete three-dimensional reconstruction of magnetic topology.

\subsubsection{2D Grad--Shafranov Reconstruction}
\label{sec:gs_reconstruction}

We further investigated the large-scale magnetic-field configuration using two-dimensional (2D) Grad--Shafranov (GS) reconstruction, a technique for recovering the cross-sectional structure of coherent magnetic configurations from single-spacecraft measurements under the assumption of approximate two-dimensional magnetohydrostatic equilibrium \citep{HauSonnerup1999,HuSonnerup2001,HuSonnerup2002}. The reconstruction was performed using magnetic-field and plasma measurements from the Wind spacecraft. Magnetic-field data from the Wind Magnetic Field Investigation (MFI) and plasma data from the Wind Solar Wind Experiment (SWE) were processed at 1-minute cadence. The de Hoffmann--Teller (HT) velocity was estimated using a least-squares approach and used to define the reconstruction frame \citep{KhrabrovSonnerup1998}. The invariant axis was estimated using minimum variance analysis (MVA) of the magnetic field, with the intermediate-variance eigenvector adopted as the candidate axis. The orientation of this axis is important because uncertainties in its determination can affect the reconstructed geometry and other structural properties \citep{HuSonnerup2002}.

In the GS framework, the cross-sectional magnetic-field structure is described by a magnetic flux function, $A$, and the axial magnetic-field component, $B_z$. Under the 2D magnetohydrostatic approximation, the transverse pressure is expressed as
\begin{equation}
P_t(A)=p+\frac{B_z^2}{2\mu_0},
\label{eq:transverse_pressure}
\end{equation}
where $p$ is the plasma thermal pressure and $\mu_0$ is the permeability of free space. Under this approximation, both the axial magnetic-field component and the transverse pressure are functions of the flux function, $A$ \citep{HuSonnerup2001,HuSonnerup2002}. The spatial distribution of the flux function is governed by the 2D Grad-Shafranov equation,
\begin{equation}
\frac{\partial^2 A}{\partial x^2}
+\frac{\partial^2 A}{\partial y^2}
=-\mu_0\frac{dP_t(A)}{dA},
\label{eq:gs_equation}
\end{equation}
where $x$ and $y$ denote coordinates in the reconstruction plane, perpendicular to the invariant axis. This equation relates the curvature of the magnetic flux function to the variation of transverse pressure with the flux function and forms the basis of the 2D GS reconstruction method \citep{HauSonnerup1999,HuSonnerup2001,HuSonnerup2002}. The observed magnetic-field and plasma profiles were transformed into the reconstruction coordinate system, and the transverse-pressure and axial-field profiles were fitted as functions of the flux function. These fitted relationships were then used in the numerical GS solver to obtain the two-dimensional distribution of the flux function and the corresponding magnetic-field configuration.

To account approximately for the influence of plasma flow, an Alfv\'enic Mach-number-based correction was incorporated into the reconstruction \citep{SonnerupEtAl2006}. The resulting 2D GS map was used to examine the cross-sectional magnetic-field structure and its coherence. The Aditya-L1 trajectory was overlaid on the Wind-derived GS reconstruction using the relative spacecraft positions transformed into the reconstruction coordinate system, enabling a geometrical comparison of the spacecraft trajectory with the reconstructed magnetic-field structure.

Successful 2D GS reconstructions yielding well-defined, closed magnetic-flux contours were obtained for ICME 2 and ICME 7, both classified as dispersion-less. As shown in Figure 3, these events exhibit coherent, flux-rope-like cross-sections with closed contours of $A$ and distinct enhancements in $B_z$. The overlaid Aditya-L1 trajectories indicate the spacecraft paths relative to the reconstructed structures. In contrast, the dispersed and mixed events did not yield comparably well-defined, coherent flux-rope-like cross-sections in the analyzed intervals. This contrast suggests an association between dispersion-less particle-onset profiles and coherent magnetic structures in the two successfully reconstructed events.

\begin{figure*}[ht!]
\centering
\includegraphics[width=\textwidth]{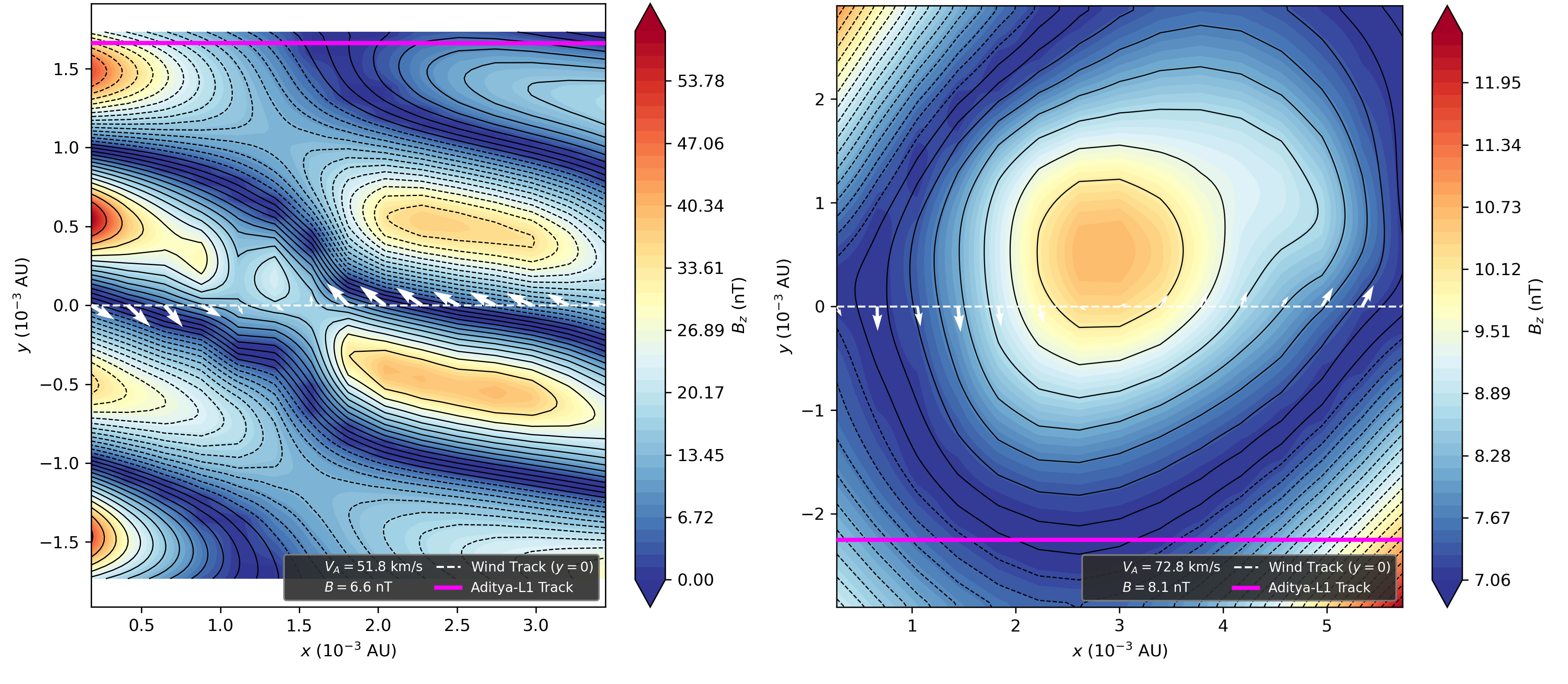}
\caption{2D Grad-Shafranov (GS) reconstructions of the cross-sectional magnetic flux function (contours) and axial magnetic field $B_z$ (color scale) for two dispersion-less ICME events: (left) ICME 2 and (right) ICME 7. White dashed lines represent the Wind spacecraft trajectory ($y = 0$), white arrows denote the transverse magnetic field vectors, and the solid magenta lines indicate the trajectory track of the Aditya-L1 spacecraft transformed into the reconstruction plane. Similar coherent, closed-contour flux-rope topologies were absent in all dispersed and mixed ICME cases.}
\label{fig:general}
\end{figure*}

\subsubsection{Minimum Variance Analysis and Magnetic-Field Hodograms}

Minimum Variance Analysis (MVA) was applied to the magnetic-field measurements from the Aditya-L1 magnetometer to characterize the directional variability of the magnetic field within the selected intervals of all ten ICMEs. The covariance matrix of the three magnetic-field components was diagonalized to obtain the eigenvalues, $\lambda_1 \geq \lambda_2 \geq \lambda_3$, and their corresponding eigenvectors, defining the maximum, intermediate, and minimum-variance directions, respectively. The magnetic-field measurements were subsequently projected onto this principal-axis frame to construct hodograms and examine the field rotation and geometrical organization. The eigenvalue ratio $\lambda_2/\lambda_3$ was used to assess the degree to which the minimum-variance direction is constrained, with values greater than approximately 3 indicating a comparatively well-defined variance plane \citep{SonnerupScheible1998}. However, this criterion alone does not establish the presence of a coherent flux rope, and the MVA results were therefore interpreted alongside the reduced magnetic helicity and Grad--Shafranov reconstruction.

The MVA results exhibit a clear variation in the degree of directional organization among the ten events. Among the seven events classified as dispersed, ICMEs 1, 3, 4, 8, 9, and 10 have $\lambda_2/\lambda_3$ ratios of 1.60, 1.85, 2.28, 1.53, 1.83, and 1.74, respectively, indicating that the minimum-variance direction is comparatively weakly constrained over the selected intervals. ICME 6 is an exception, with $\lambda_2/\lambda_3 = 3.44$, suggesting a better-defined variance plane despite its dispersed particle-onset classification. The remaining dispersed event, ICME 4, has an intermediate ratio of 2.28, which indicates moderate rather than strong separation between the intermediate and minimum-variance eigenvalues. Taken together, these results show that the dispersed events do not exhibit a uniform magnetic-field geometry: several have relatively weak variance-axis separation, whereas ICME 6 satisfies the adopted eigenvalue-ratio criterion.

\begin{figure*}[ht!]
\centering
\includegraphics[width=\textwidth]{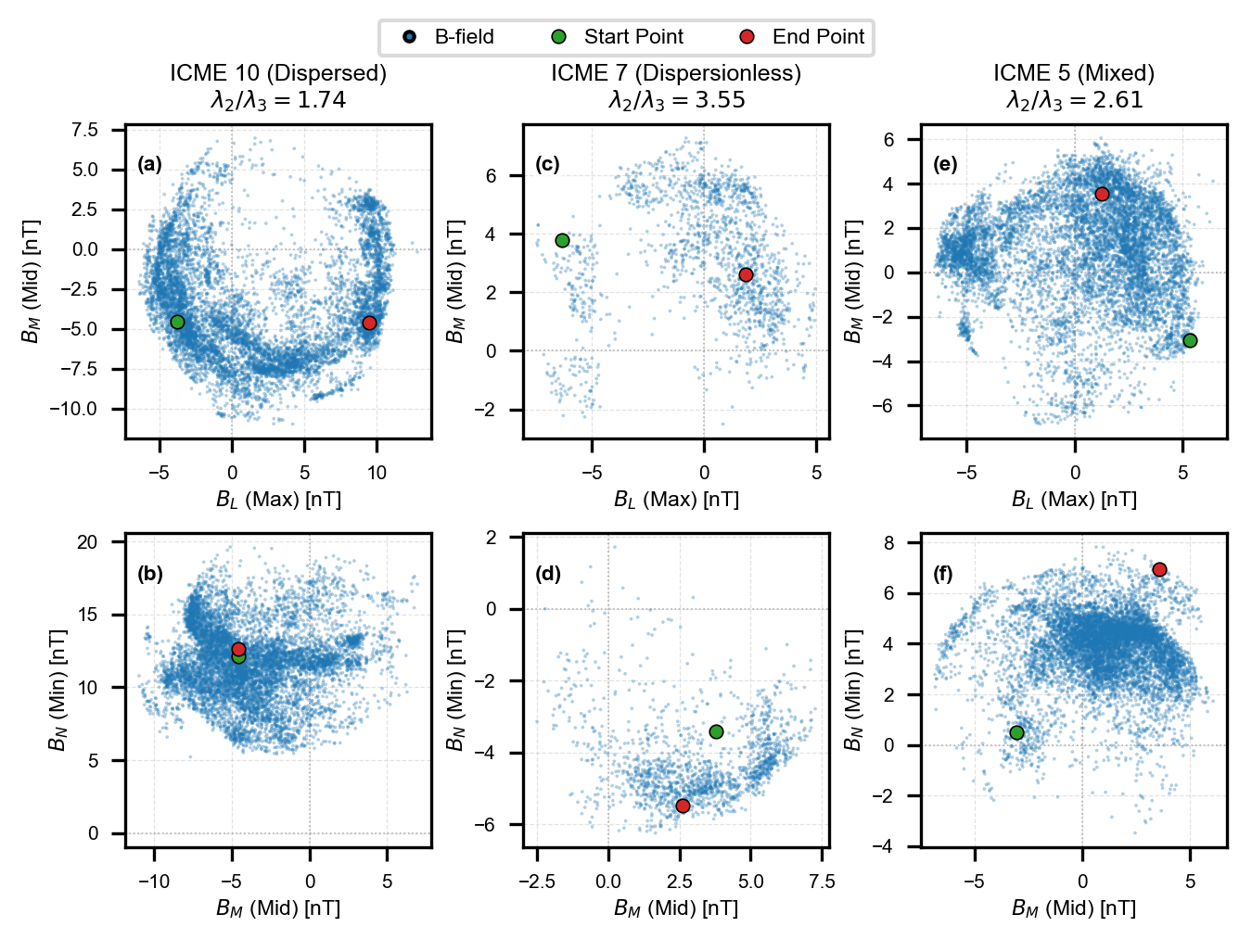}
\caption{Representative minimum variance analysis (MVA) magnetic-field hodograms projected onto the principal-axis frame ($L, M, N$ corresponding to maximum, intermediate, and minimum-variance directions) for three selected ICMEs observed by Aditya-L1: (a)–(b) ICME 10 (dispersed class, $\lambda_2/\lambda_3 = 1.74$), (c)–(d) ICME 7 (dispersionless class, $\lambda_2/\lambda_3 = 3.55$), and (e)–(f) ICME 5 (mixed class, $\lambda_2/\lambda_3 = 2.61$). The top row ((a), (c), (e)) presents the $B_M$ vs. $B_L$ hodogram, while the bottom row ((b), (d), (f)) presents the $B_N$ vs. $B_M$ hodogram. Blue points represent individual magnetic field measurements, with the green and red circles marking the start and end points of the analysis interval, respectively.}
\label{fig:general}
\end{figure*}

The two dispersionless events exhibit comparatively higher eigenvalue ratios. ICME 2 has $\lambda_2/\lambda_3 = 3.42$, while ICME 7 has $\lambda_2/\lambda_3 = 3.55$. These values indicate a better-defined minimum-variance direction than that found in most of the dispersed events. Such directional organization may be consistent with a more coherent magnetic-field structure over the analyzed intervals. Nevertheless, the eigenvalue ratios do not independently demonstrate that the field forms a flux rope or that particle transport occurs with limited scattering. 

ICME 5, classified as mixed, has an intermediate eigenvalue ratio of $\lambda_2/\lambda_3 = 2.61$. This value indicates a greater separation between the intermediate and minimum-variance eigenvalues than in most of the dispersed events, but it does not meet the adopted threshold of 3. The result is compatible with an intermediate degree of directional organization, although the eigenvalue ratio alone cannot establish a connection between the mixed particle-onset behavior and the magnetic topology.

Overall, seven of the ten events have $\lambda_2/\lambda_3 < 3$, whereas three events-ICMEs 2, 6, and 7-exceed this threshold. The occurrence of a comparatively well-defined variance plane in ICME 6, despite its dispersed classification, also demonstrates that the particle-onset classes cannot be mapped directly onto a single MVA criterion. The hodograms and principal-axis projections therefore provide a complementary geometrical diagnostic rather than a standalone classification of magnetic topology. Their interpretation in conjunction with reduced magnetic helicity and GS reconstruction provides a more comprehensive assessment of the magnetic-field organization within the selected ICME intervals. 

Figure 4 presents the MVA hodograms for three representative ICMEs corresponding to the dispersed, dispersionless, and mixed particle-onset classes illustrated in Figures~1 and 2. The hodograms show the magnetic-field projections onto the principal-variance directions and provide a geometrical view of the field variability and rotation within the selected intervals. These representative cases facilitate a qualitative comparison of the magnetic-field configurations associated with the three particle-onset classes.

\subsection{Estimation of ICME Shock Parameters}
\label{subsec:shock_parameters}

The shock parameters of the identified ICME-driven events were estimated using magnetic-field measurements from the Aditya-L1 magnetometer and proton plasma measurements from the Solar Wind Ion Spectrometer (SWIS) onboard Aditya-L1. For each event, upstream and downstream intervals were selected on either side of the identified shock time, excluding the immediate shock transition. To assess the sensitivity of the derived parameters to the choice of these intervals, the averaging-window durations were systematically varied from 6 to 24 minutes in 3-minute increments. The mean magnetic-field and plasma parameters obtained for each upstream--downstream interval pair were used to calculate the shock parameters, and the resulting solutions were combined to obtain robust estimates, following the ensemble-based approach described by \citet{Trotta2022,Trotta2024}.

The shock normal was determined using the mixed-mode 3 (MX3) method, which combines magnetic-field and plasma-velocity measurements across the shock \citep{AbrahamShrauner1972,AbrahamShraunerYun1976,Schwartz1998,PaschmannSchwartz2000,Trotta2022}. The unit shock-normal vector is given by
\begin{equation}
\hat{\mathbf{n}}_{\mathrm{MX3}} =
\frac{
(\Delta\mathbf{B}\times\Delta\mathbf{V})
\times\Delta\mathbf{B}
}{
\left|
(\Delta\mathbf{B}\times\Delta\mathbf{V})
\times\Delta\mathbf{B}
\right|
},
\label{eq:mx3_normal}
\end{equation}
where $\Delta\mathbf{B}=\mathbf{B}_d-\mathbf{B}_u$ and
$\Delta\mathbf{V}=\mathbf{V}_d-\mathbf{V}_u$, with subscripts $u$
and $d$ denoting the upstream and downstream states, respectively.
The sign of the normal was chosen such that its GSE $x$ component
is positive. The angle between the upstream magnetic field and the
shock normal, which characterizes the shock geometry, was calculated
as
\begin{equation}
\theta_{Bn} =
\cos^{-1}\left(
\frac{\left|\mathbf{B}_u\cdot\hat{\mathbf{n}}_{\mathrm{MX3}}\right|}
{|\mathbf{B}_u|}
\right).
\label{eq:theta_bn}
\end{equation}
This definition gives $0^\circ \leq \theta_{Bn} \leq 90^\circ$,
with larger angles corresponding to more quasi-perpendicular
shock geometries.

The magnetic-field and proton-density compression ratios were
calculated as
\begin{equation}
r_B=\frac{|\mathbf{B}_d|}{|\mathbf{B}_u|},
\qquad
r_n=\frac{n_d}{n_u},
\label{eq:compression_ratios}
\end{equation}
where $n$ denotes the proton number density. The shock propagation
speed in the spacecraft frame was initially estimated from mass
conservation across the shock \citep{Trotta2022}:
\begin{equation}
V_s =
\left|
\frac{
n_d(\mathbf{V}_d\cdot\hat{\mathbf{n}}_{\mathrm{MX3}})
-
n_u(\mathbf{V}_u\cdot\hat{\mathbf{n}}_{\mathrm{MX3}})
}{
n_d-n_u
}
\right|.
\label{eq:shock_speed}
\end{equation}
For weak density compressions ($r_n<1.30$), or when the mass-continuity estimate was nonphysical, an alternative estimate was obtained using the upstream normal-flow speed and an approximate fast-magnetosonic Mach number. The latter was estimated from the density compression ratio using the gas-dynamic shock relation for an adiabatic index of $\gamma=5/3$ \citep{LandauLifshitz1959,GoedbloedPoedts2004}:
\begin{equation}
M_{\mathrm{ms}}^{\mathrm{est}} =
\sqrt{
\max\left[
1,\,
\frac{2r_n}
{(\gamma+1)-(\gamma-1)r_n}
\right]
},
\label{eq:fallback_mach}
\end{equation}
provided that the denominator is positive. The corresponding fallback shock speed was calculated as
\begin{equation}
V_s =
\left|\mathbf{V}_u\cdot\hat{\mathbf{n}}_{\mathrm{MX3}}\right|
+
M_{\mathrm{ms}}^{\mathrm{est}}V_f,
\label{eq:fallback_shock_speed}
\end{equation}
where $V_f$ is the upstream fast-mode magnetosonic speed. This fallback was applied at the individual averaging-window level when the mass-continuity estimate was unreliable. The final shock parameters were then obtained from the ensemble of valid upstream–downstream window combinations. This fallback provides an approximate shock-speed estimate when the mass-continuity method is unreliable; it does not represent a complete solution of the MHD Rankine–Hugoniot relations.

The upstream Alfv\'en speed was calculated from the upstream magnetic-field magnitude and mass density using the standard magnetohydrodynamic expression \citep{GoedbloedPoedts2004}:
\begin{equation}
V_A=\frac{|\mathbf{B}_u|}{\sqrt{\mu_0\rho_u}},
\label{eq:alfven_speed}
\end{equation}
where $\mu_0$ is the vacuum permeability and $\rho_u$ is the upstream mass density. For a proton-dominated plasma, $\rho_u\simeq m_p n_u$, where $m_p$ is the proton mass. The upstream sound speed, $c_s$, was estimated from the proton thermal parameter. The fast-mode magnetosonic speed was then calculated using the standard MHD dispersion relation \citep{Priest2014,GoedbloedPoedts2004}:
\begin{equation}
V_f =
\left\{
\frac{1}{2}
\left[
V_A^2+c_s^2+
\sqrt{
(V_A^2+c_s^2)^2
-4V_A^2c_s^2\cos^2\theta_{Bn}
}
\right]
\right\}^{1/2},
\label{eq:fast_mode_speed}
\end{equation}
where $\theta_{Bn}$ is the angle between the upstream magnetic field and the shock normal.

The Alfv\'enic and fast-mode Mach numbers were estimated from the shock speed relative to the upstream normal plasma flow. The corresponding relative normal speed was defined as
\begin{equation}
V_{\mathrm{rel},n}
=
V_s-\left|\mathbf{V}_u\cdot\hat{\mathbf{n}}_{\mathrm{MX3}}\right|,
\label{eq:relative_normal_speed}
\end{equation}
giving
\begin{equation}
M_A=\frac{V_{\mathrm{rel},n}}{V_A},
\qquad
M_{\mathrm{ms}}=\frac{V_{\mathrm{rel},n}}{V_f}.
\label{eq:mach_numbers}
\end{equation}

The radial shock speed was estimated from the shock-normal speed and the GSE $x$ component of the unit normal as
\begin{equation}
V_{s,x}=\frac{V_s}{|n_x|},
\label{eq:radial_shock_speed}
\end{equation}
where $n_x$ is the GSE $x$ component of $\hat{\mathbf{n}}_{\mathrm{MX3}}$. The final shock-normal components and scalar parameters were obtained from the median of the valid solutions across the different averaging-window combinations. The derived shock-normal components, $\theta_{Bn}$, shock radial speed, compression ratios, and Mach numbers are summarized in Table~\ref{tab:shock_parameters}.

\begin{deluxetable*}{clcccccccc}
\tablecaption{Summary of the derived shock-normal orientations, radial shock speeds, compression ratios, and Mach numbers for the ICMEs observed by Aditya-L1.\label{tab:shock_parameters}}
\tablehead{
\colhead{ICME No.} &
\colhead{Shock Time} &
\colhead{Type of Flux} &
\colhead{$(n_x,n_y,n_z)$} &
\colhead{$\theta_{Bn}$} &
\colhead{$V_{s,x}$} &
\colhead{$r_n$} &
\colhead{$r_B$} &
\colhead{$M_A$} &
\colhead{$M_{\rm ms}$} \\
&
&
\colhead{Variations} &
&
\colhead{(deg)} &
\colhead{(km s$^{-1}$)} &
&
&
&
}
\startdata
1  & 2024-02-11 01:28 & Dispersed     & $(0.263,\,0.376,\,0.889)$    & 51.91 & 802.55  & 1.00 & 1.46 & 1.03  & 1.00 \\
2  & 2024-03-24 14:10 & Dispersionless & $(0.918,\,0.142,\,0.371)$   & 51.92 & 903.49  & 2.93 & 1.62 & 4.27  & 2.70 \\
3  & 2024-07-23 20:00 & Dispersed     & $(0.969,\,0.246,\,0.005)$    & 17.46 & 497.49  & 1.21 & 1.01 & 1.16  & 1.14 \\
4  & 2024-09-17 10:08 & Dispersed     & $(0.440,\,-0.897,\,-0.037)$  & 23.34 & 872.16  & 1.18 & 1.20 & 1.18  & 1.12 \\
5  & 2024-10-10 14:49 & Mixed         & $(0.777,\,-0.600,\,-0.189)$  & 56.69 & 934.93  & 3.89 & 1.62 & 2.09  & 1.68 \\
6  & 2025-02-26 02:47 & Dispersed     & $(0.201,\,-0.955,\,0.216)$   & 41.61 & 720.71  & 1.03 & 1.02 & 1.35  & 1.02 \\
7  & 2025-06-01 05:19 & Dispersionless & $(0.947,\,-0.309,\,0.083)$  & 31.39 & 1182.80 & 3.47 & 2.70 & 10.42 & 4.90 \\
8  & 2025-09-01 20:28 & Dispersed     & $(0.809,\,0.587,\,0.044)$    & 58.70 & 822.87  & 1.63 & 1.45 & 3.02  & 2.35 \\
9  & 2025-11-12 00:00 & Dispersed     & $(0.823,\,-0.009,\,-0.567)$  & 61.91 & 841.88  & 1.08 & 1.79 & 1.27  & 1.05 \\
10 & 2026-01-10 16:03 & Dispersed     & $(0.377,\,0.394,\,0.838)$    & 48.78 & 908.08  & 1.02 & 0.96 & 1.02  & 1.01 \\
\enddata
\end{deluxetable*}

\subsection{Estimation of Energetic Ion Transport Parameters}
\label{subsec:qlt_transport}

To investigate the possible influence of magnetic turbulence on energetic-particle transport during the selected ICMEs, quasi-linear theory (QLT) was used to estimate the pitch-angle diffusion coefficient, parallel mean free path, and parallel spatial diffusion coefficient \citep{Jokipii1966,Schlickeiser2002}. Magnetic-field measurements from the Aditya-L1 magnetometer were analyzed together with the solar-wind speed obtained from the ASPEX-SWIS proton bulk data. The transport parameters were calculated within one-hour magnetic-field windows advanced in 30-minute increments. This overlapping-window approach provides a time-resolved characterization of the magnetic fluctuations while retaining sufficient data for spectral estimation. For each ICME, the final transport parameters were obtained as the median of the window-level estimates.

For each interval, the mean magnetic-field vector was calculated as
\begin{equation}
\mathbf{B}_0 = \langle \mathbf{B}(t) \rangle,
\qquad
B_0 = |\mathbf{B}_0|,
\label{eq:qlt_mean_field}
\end{equation}
and the magnetic fluctuations were defined as
\begin{equation}
\delta\mathbf{B}(t)=\mathbf{B}(t)-\mathbf{B}_0.
\end{equation}
The fluctuations perpendicular to the mean magnetic field were obtained from
\begin{equation}
\delta\mathbf{B}_{\perp}
=
\delta\mathbf{B}
-
(\delta\mathbf{B}\cdot\hat{\mathbf{b}}_0)
\hat{\mathbf{b}}_0,
\end{equation}
where $\hat{\mathbf{b}}_0=\mathbf{B}_0/B_0$. The perpendicular magnetic-fluctuation power spectral density (PSD), $P_{\perp}(f)$, was estimated using Welch's method \citep{Welch1967} by summing the PSDs of the three components of $\delta\mathbf{B}_{\perp}$. The magnetic-field fluctuation spectra measured in the spacecraft frame were used to estimate the resonant frequencies under the Taylor frozen-in hypothesis, which assumes that the measured fluctuations are predominantly advected past the spacecraft by the solar-wind flow \citep{Taylor1938,Treumann2019}.

For a proton with kinetic energy $E$, the particle speed was calculated using the nonrelativistic expression
\begin{equation}
v=\sqrt{\frac{2E}{m_p}},
\label{eq:qlt_particle_speed}
\end{equation}
where $m_p$ is the proton mass. The proton gyrofrequency was defined as
\begin{equation}
\Omega_p=\frac{eB_0}{m_p},
\label{eq:qlt_gyrofrequency}
\end{equation}
where $e$ is the elementary charge. Under the fundamental cyclotron resonance approximation, the resonant spacecraft-frame frequency for a particle with pitch-angle cosine $\mu$ is
\begin{equation}
f_{\mathrm{res}}(\mu)
=
\left|\frac{V_{\mathrm{sw}}}{v\mu}\right|
\frac{\Omega_p}{2\pi},
\label{eq:qlt_resonance}
\end{equation}
where $V_{\mathrm{sw}}$ is the solar-wind speed. The resonant PSD was obtained by evaluating $P_{\perp}(f)$ at $f_{\mathrm{res}}$, and the pitch-angle diffusion coefficient was calculated as
\begin{equation}
D_{\mu\mu}(\mu)
=
\frac{\pi\Omega_p}{2B_0^2}
(1-\mu^2)
f_{\mathrm{res}}(\mu)
P_{\perp}\!\left[f_{\mathrm{res}}(\mu)\right].
\label{eq:qlt_dmm}
\end{equation}

The parallel mean free path was subsequently estimated from
\begin{equation}
\lambda_{\parallel}
=
\frac{3v}{4}
\int_{0.05}^{0.95}
\frac{(1-\mu^2)^2}{D_{\mu\mu}(\mu)}
\,d\mu,
\label{eq:qlt_mfp}
\end{equation}
assuming symmetric pitch-angle scattering about $\mu=0$. The integration domain was restricted to $0.05\leq|\mu|\leq0.95$ to avoid numerical difficulties associated with the vicinity of $90^\circ$ pitch angle, where the resonant frequency becomes very large, and the extreme pitch-angle boundaries, where the pitch-angle diffusion coefficient approaches zero. This restriction provides a numerically stable integration domain and is consistent with the treatment adopted in previous studies \citep{Giacalone2020}. A numerical floor of $D_{\mu\mu}=10^{-12}\,\mathrm{s}^{-1}$ was imposed where necessary to prevent numerical singularities during integration. The resulting mean free path was converted to astronomical units (AU). Finally, the parallel spatial diffusion coefficient was obtained from
\begin{equation}
\kappa_{\parallel}
=
\frac{1}{3}v\lambda_{\parallel},
\label{eq:qlt_kappa}
\end{equation}
and expressed in $\mathrm{cm^2\,s^{-1}}$.

The applicability of these estimates depends on the frequency coverage of the measured magnetic-field spectra and the underlying turbulence properties. It is well established that energetic-particle diffusion coefficients depend sensitively on the turbulence geometry, including the partitioning between slab and two-dimensional (2D) fluctuations, turbulence anisotropy, and spectral break scales \citep{Zhao2017,Zhao2018}. In this study, standard QLT is adopted as a first-order diagnostic to estimate $D_{\mu\mu}(\mu)$ and $\lambda_{\parallel}$ from the observed transverse magnetic-field PSD, $P_{\perp}(f)$.

With the native sampling cadence of $\Delta t=10\,\mathrm{s}$ for the Aditya-L1 magnetometer, the Nyquist frequency is
\begin{equation}
f_{\mathrm{Nyq}}=\frac{1}{2\Delta t}=0.05\,\mathrm{Hz}.
\end{equation}
For the one-hour Welch spectral windows, using 256-sample segments
at a 10-s cadence, the frequency-bin spacing is
$\Delta f\simeq3.9\times10^{-4}\,\mathrm{Hz}$.
The upper frequency limit is set by the Nyquist frequency,
$f_{\rm Nyquist}=1/(2\Delta t)=0.05\,\mathrm{Hz}$.
Therefore, the measured spectra span approximately
$3.9\times10^{-4}\leq f\leq0.05\,\mathrm{Hz}$. Under the Taylor frozen-in hypothesis and the fundamental cyclotron resonance condition, $k_{\parallel}v|\mu|=\Omega_p$, the spacecraft-frame resonant frequency is given by Equation 23. Requiring this frequency to remain below the Nyquist frequency defines the minimum pitch-angle cosine for which the resonance is directly resolved:
\begin{equation}
|\mu|_{\min}
=
\frac{V_{\mathrm{sw}}}{v}
\frac{\Omega_p}{2\pi f_{\mathrm{Nyq}}}.
\label{eq:qlt_mu_min}
\end{equation}
The proton gyrofrequency in cyclic frequency units is $\Omega_p/(2\pi)=eB_0/(2\pi m_p)\simeq0.01524 B_0[\mathrm{nT}]\,\mathrm{Hz}$. Across the ten ICME events, $B_0$ ranges from approximately $4.85$ to $17.45\,\mathrm{nT}$, corresponding to $\Omega_p/(2\pi)\simeq0.074$--$0.266\,\mathrm{Hz}$, while $V_{\mathrm{sw}}$ (statistical median of solar wind speed measurements falling within that specific event window to get a representative average bulk speed) ranges from approximately $279$ to $652\,\mathrm{km\,s^{-1}}$.

For $1.2\,\mathrm{MeV}$ protons, with $v\simeq1.52\times10^4\,\mathrm{km\,s^{-1}}$, the estimated $|\mu|_{\min}$ ranges from approximately 0.04 to 0.19, with a median of about 0.10. Thus, the resonant frequencies are directly resolved below the Nyquist frequency over nearly the entire pitch-angle domain, except close to $90^\circ$. For $150\,\mathrm{keV}$ protons, with $v\simeq5.36\times10^3\,\mathrm{km\,s^{-1}}$, $|\mu|_{\min}$ ranges from approximately 0.11 to 0.54 across the event sample. For representative disturbed events, the corresponding threshold is typically $\mu\gtrsim0.25$--$0.40$, equivalent to pitch angles of approximately $66^\circ$--$75^\circ$. Consequently, the 10-s magnetometer measurements directly resolve the resonant power for moderate-to-large pitch-angle cosines, whereas particles with pitch angles approaching $90^\circ$ require resonant frequencies above the Nyquist limit.

As $\mu\rightarrow0$, corresponding to pitch angles approaching $90^\circ$, the resonant frequency exceeds the Nyquist frequency and enters the unresolved high-frequency regime. This is also associated with the classical $90^\circ$ resonance gap in QLT, for which $D_{\mu\mu}$ approaches zero. To maintain numerical stability, the integration domain was therefore restricted to $0.05\leq\mu\leq0.95$, which is arbitrary but logical choice, and the numerical floor described above was applied to $D_{\mu\mu}$. For resonant frequencies exceeding the Nyquist frequency, the PSD was approximated by its value at the Nyquist boundary, i.e.,
\begin{equation}
P_{\perp}(f_{\mathrm{res}}>f_{\mathrm{Nyq}})
=
P_{\perp}(f_{\mathrm{Nyq}}).
\label{eq:qlt_psd_clamp}
\end{equation}
This endpoint prescription provides a numerical approximation for evaluating the diffusion coefficient where the resonant power is not directly measured. Since solar-wind magnetic spectra generally steepen at frequencies approaching and exceeding the ion gyroscale, this prescription may overestimate the actual high-frequency power and, consequently, the scattering rate. It may therefore lead to an underestimate of $\lambda_{\parallel}$ in pitch-angle regions where the resonance lies beyond the measured frequency range. The resulting estimates should accordingly be interpreted with this frequency-coverage limitation in mind.

The measured magnetic fluctuation amplitudes further indicate that the weak-turbulence assumption underlying classical QLT is not uniformly satisfied across the event sample. The event-median normalized fluctuation amplitudes, $\delta B_{\mathrm{rms}}/B_0$, range from approximately 0.11 to 0.52 (median $\sim0.27$) for the dispersed events, whereas values of approximately 0.55--0.66 are obtained in the strongly disturbed upstream regions of the dispersionless and mixed events. Classical QLT formally assumes weak turbulence, $\delta B/B_0\ll1$, and does not fully account for nonlinear resonance broadening or cross-field transport. More advanced transport descriptions incorporate additional effects that may become important under strongly turbulent conditions \citep[e.g.,][]{Subashchandar2025}. Nevertheless, applying QLT consistently across the event sample provides a standardized, first-order diagnostic for comparing the relative particle-scattering conditions among the different event classes. The resulting mean free paths, particularly the small values obtained during strongly turbulent intervals (approximately $0.0037$-$0.0099\,\mathrm{AU}$), should therefore be interpreted as classical QLT-dependent estimates rather than direct observational measurements of the particle scattering length. The present analysis does not incorporate nonlinear guiding-center transport or other nonlinear corrections to particle diffusion. 

The transport parameters obtained for each ICME event are summarized in Table~\ref{tab:qlt_transport}. The table presents the mean magnetic-field strength ($B_0$), normalized magnetic fluctuation amplitude ($\delta B/B_0$), parallel mean free path ($\lambda_{\parallel}$), parallel spatial diffusion coefficient ($\kappa_{\parallel}$), and mean pitch-angle diffusion coefficient ($D_{\mu\mu}$) for protons at 150 keV and 1.2 MeV. These parameters provide a comparative assessment of the magnetic turbulence and particle-scattering environment associated with the dispersed, dispersionless, and mixed event classes.

\begin{deluxetable*}{cccccccccccc}
\tabletypesize{\scriptsize}
\tablewidth{0pt}
\tablecolumns{12}
\tablecaption{Quasi-linear transport parameters and magnetic turbulence levels derived for the ten ICMEs observed by Aditya-L1. \label{tab:qlt_transport}}

\tablehead{
\colhead{ICME} & \colhead{Start Time} & \colhead{End Time} & \colhead{Type of flux} &
\colhead{$B_0$} & \colhead{$\delta B_{\rm rms}/B_0$} &
\multicolumn{3}{c}{150 keV} & \multicolumn{3}{c}{1.2 MeV} \\
\cline{7-9}\cline{10-12}
\colhead{No.} & \colhead{} & \colhead{} & \colhead{variations} &
\colhead{(nT)} & \colhead{} &
\colhead{$\lambda_\parallel$ (AU)} & \colhead{$\kappa_\parallel$ (cm$^2$ s$^{-1}$)} & \colhead{$\langle D_{\mu\mu}\rangle$ (s$^{-1}$)} &
\colhead{$\lambda_\parallel$ (AU)} & \colhead{$\kappa_\parallel$ (cm$^2$ s$^{-1}$)} & \colhead{$\langle D_{\mu\mu}\rangle$ (s$^{-1}$)}
}

\startdata
1  & 2024-02-06 13:23 & 2024-02-11 01:28 & Dispersed       & 7.20  & 0.236 & 0.0555 & $1.48\times10^{20}$ & $4.77\times10^{-4}$ & 0.0894 & $6.76\times10^{20}$ & $8.38\times10^{-4}$ \\
2  & 2024-03-24 00:00 & 2024-03-24 14:10 & Dispersion-less & 9.31  & 0.584 & 0.0084 & $2.25\times10^{19}$ & $3.87\times10^{-3}$ & 0.0099 & $7.45\times10^{19}$ & $9.02\times10^{-3}$ \\
3  & 2024-07-22 16:07 & 2024-07-23 20:00 & Dispersed       & 5.39  & 0.270 & 0.0230 & $6.15\times10^{19}$ & $9.52\times10^{-4}$ & 0.0303 & $2.29\times10^{20}$ & $1.76\times10^{-3}$ \\
4  & 2024-09-17 03:41 & 2024-09-17 10:08 & Dispersed       & 17.45 & 0.233 & 0.0131 & $3.51\times10^{19}$ & $1.36\times10^{-3}$ & 0.0230 & $1.74\times10^{20}$ & $2.03\times10^{-3}$ \\
5  & 2024-10-09 12:00 & 2024-10-10 14:49 & Mixed           & 4.85  & 0.552 & 0.0049 & $1.30\times10^{19}$ & $4.85\times10^{-3}$ & 0.0064 & $4.85\times10^{19}$ & $9.99\times10^{-3}$ \\
6  & 2025-02-25 08:28 & 2025-02-26 02:47 & Dispersed       & 8.98  & 0.387 & 0.0129 & $3.45\times10^{19}$ & $1.54\times10^{-3}$ & 0.0252 & $1.91\times10^{20}$ & $2.63\times10^{-3}$ \\
7  & 2025-05-31 13:00 & 2025-06-01 05:19 & Dispersion-less & 5.15  & 0.658 & 0.0037 & $9.92\times10^{18}$ & $7.36\times10^{-3}$ & 0.0046 & $3.48\times10^{19}$ & $1.38\times10^{-2}$ \\
8  & 2025-08-31 18:34 & 2025-09-01 20:28 & Dispersed       & 5.51  & 0.516 & 0.0118 & $3.15\times10^{19}$ & $2.82\times10^{-3}$ & 0.0128 & $9.70\times10^{19}$ & $5.14\times10^{-3}$ \\
9  & 2025-11-11 06:00 & 2025-11-12 00:00 & Dispersed       & 7.21  & 0.108 & 0.0474 & $1.27\times10^{20}$ & $5.90\times10^{-4}$ & 0.0793 & $5.99\times10^{20}$ & $8.70\times10^{-4}$ \\
10 & 2026-01-09 14:45 & 2026-01-10 16:03 & Dispersed       & 13.14 & 0.289 & 0.0142 & $3.80\times10^{19}$ & $1.55\times10^{-3}$ & 0.0210 & $1.59\times10^{20}$ & $3.03\times10^{-3}$ \\
\enddata
\end{deluxetable*}

\section{Results and Discussion} \label{sec:ResultsDiscussion}
\subsection{Dispersion-less Upstream Energetic Ion Events} 
To investigate the dispersion-less flux variations in the upstream regions of ICME-driven shocks, we examine two dispersion-less energetic ion enhancement events observed by the ASPEX-STEPS instrument onboard Aditya-L1: ICME 2 (observed between 00:00 UTC and 14:10 UTC on 24 March 2024; left column) and ICME 7 (observed between 13:00 UTC and 05:19 UTC on 31 May--1 June 2025; right column) as shown in Figure 5. The shaded intervals indicate the upstream energetic ion enhancement regions, with the second vertical dashed line marking the arrival of the respective ICME-driven shock at L1. In Figure 5, panels (a)-(d) show the energy-time spectrograms of energetic ion fluxes measured by the PS, IM, EP, and NP detectors of ASPEX-STEPS, respectively. Overplotted dashed black curves represent the corresponding ion pitch angles derived using the local interplanetary magnetic field measured by the Aditya-L1 MAG instrument. In both events, energetic ion flux enhancements are observed nearly simultaneously across all energy channels, without any clear energy-dependent dispersion, confirming their dispersion-less nature. 

During the shaded intervals, the derived pitch-angle distributions exhibit predominantly field-aligned streaming and bi-directional features, quantified using the first-order anisotropy index ($A$) shown in panel (e). For the 24 March 2024 event (ICME 2), the anisotropy index remains predominantly positive, indicating streaming along the magnetic field direction \citep{Jokipii1966, Sanderson1985, Ding2025}, whereas the 31 May--1 June 2025 event (ICME 7) exhibits repeated sign reversals about zero, indicative of bidirectional streaming and magnetic trapping within an upstream structure \citep{Ding2025}. 

To determine whether these streaming characteristics are associated with coherent magnetic topologies, we examine the local magnetic structure using minimum variance analysis (MVA) and reduced magnetic helicity. Both dispersionless events exhibit extended intervals of enhanced reduced magnetic helicity, with $|\sigma_m|\gtrsim0.6$ over timescales exceeding $10^3$ min as shown in panel (g), consistent with organized magnetic fluctuations \citep{Zhao2020,Zhao2021a,Zhao2021b}. MVA yields eigenvalue ratios of $\lambda_2/\lambda_3=3.42$ for ICME~2 and $\lambda_2/\lambda_3=3.55$ for ICME~7, indicating comparatively well-separated intermediate and minimum-variance eigenvalues. Although these ratios alone do not establish flux-rope topology, they provide complementary evidence of magnetic-field organization.

The magnetic structure is further examined using two-dimensional (2D) Grad--Shafranov (GS) reconstruction (Section 3.4.3), which yields well-defined closed magnetic-flux contours and distinct core-field enhancements for both ICME~2 and ICME~7. Comparable, well-defined closed-flux-contour structures were not obtained for the dispersed and mixed events in the analyzed intervals. These coherent structures may act as magnetic conduits that provide efficient field-aligned transport while suppressing cross-field diffusion. 

\begin{figure*}[!ht]
\centering
\includegraphics[width=\textwidth]{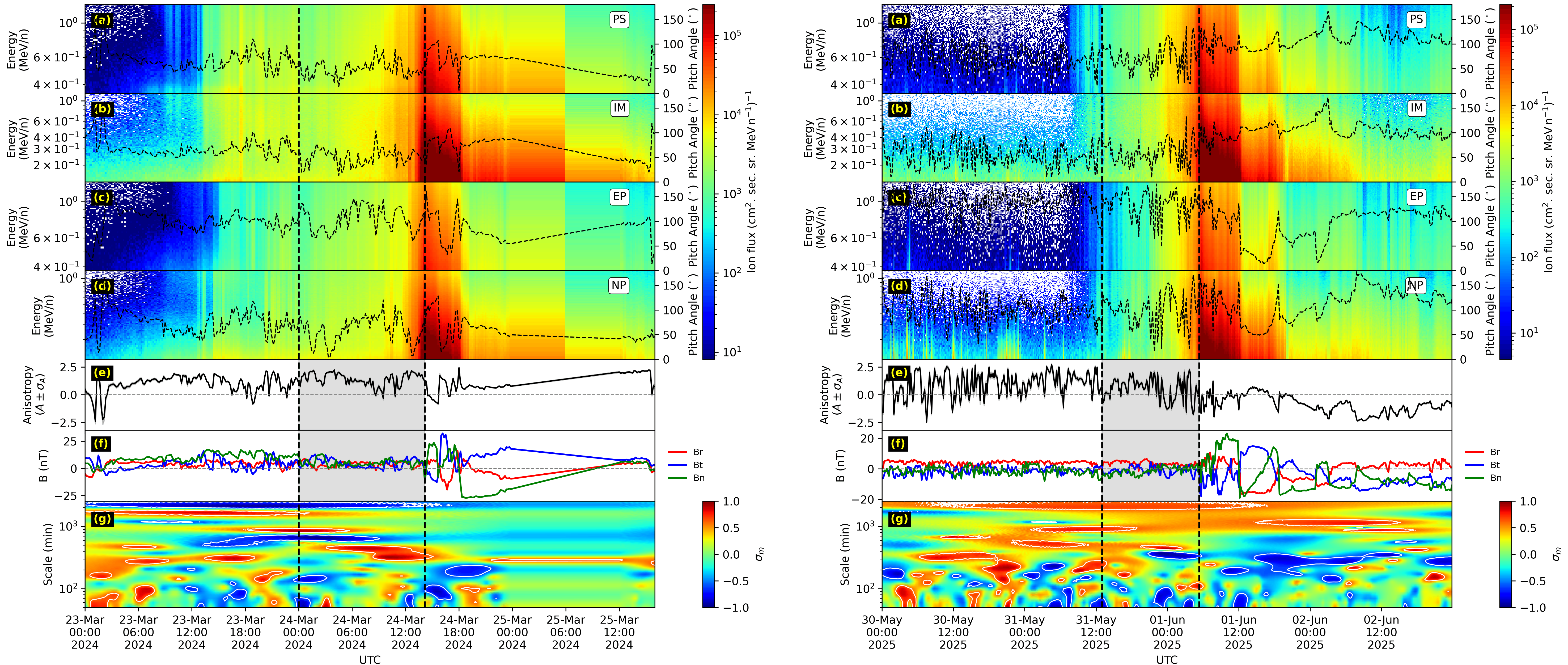}
\caption{This figure represents detailed observations of two dispersion-less upstream energetic ion enhancement events associated with ICME-driven shocks observed on 24 March 2024 (left panels; upstream enhancement interval from 00:00 UTC to 14:10 UTC) and 31 May--1 June 2025 (right panels; upstream enhancement interval from 13:00 UTC on 31 May to 05:19 UTC on 1 June). Panels (a)--(d) show energy-time spectrograms of energetic ion fluxes measured by the PS, IM, EP, and NP detectors of ASPEX-STEPS onboard Aditya-L1, respectively, with overplotted dashed black curves representing the corresponding ion pitch angles derived from Aditya-L1 MAG observations. The shaded regions indicate the upstream energetic ion enhancement intervals extending up to the shock arrival time marked by the second vertical dashed line. Panel (e) shows the temporal variation of the first-order anisotropy index ($A$), while panel (f) presents the RTN magnetic field components measured by MAG. Panel (g) displays the scale- and time-dependent normalized reduced magnetic helicity ($\sigma_m$) derived using complex Morlet wavelet analysis, where regions enclosed by white contours indicate coherent helical magnetic structures ($|\sigma_m| \gtrsim 0.6$).}
\label{fig:general}
\label{fig:general} 
\end{figure*}

The quasi-linear transport parameters provide further insight into the dispersionless energetic-ion enhancements (Table 3). For ICME~2, the parallel mean free path and spatial diffusion coefficient are $\lambda_\parallel=0.0084$~AU and $\kappa_\parallel=2.25\times10^{19}$~cm$^2$~s$^{-1}$ at 150~keV, increasing to $\lambda_\parallel=0.0099$~AU and $\kappa_\parallel=7.45\times10^{19}$~cm$^2$~s$^{-1}$ at 1.2~MeV. For ICME~7, the corresponding values are $\lambda_\parallel=0.0037$~AU and $\kappa_\parallel=9.92\times10^{18}$~cm$^2$~s$^{-1}$ at 150~keV, and $\lambda_\parallel=0.0046$~AU and $\kappa_\parallel=3.48\times10^{19}$~cm$^2$~s$^{-1}$ at 1.2~MeV. The corresponding pitch-angle diffusion coefficients, $\langle D_{\mu\mu}\rangle$, are $3.87\times10^{-3}$ and $9.02\times10^{-3}$~s$^{-1}$ for ICME~2, and $7.36\times10^{-3}$ and $1.38\times10^{-2}$~s$^{-1}$ for ICME~7, respectively. Compared with the other eight events, both ICMEs exhibit the lowest parallel mean free paths and spatial diffusion coefficients and the highest pitch-angle diffusion coefficients at both energies. Within the adopted transport model, these values indicate stronger effective pitch-angle scattering and reduced spatial transport along the magnetic field. Combined with the extended intervals of enhanced reduced magnetic helicity ($|\sigma_m|\gtrsim0.6$) and the coherent magnetic structures identified through MVA and 2D Grad--Shafranov reconstruction, these results are consistent with particle confinement within an organized magnetic structure. Such confinement may limit particle escape and contribute to the nearly simultaneous energetic-ion enhancements across energy channels, although the onset timing may also depend on the injection geometry and temporal evolution.

The shock parameters provide additional context for these enhancements. ICME~2 has a radial shock speed of $V_{s,x}=903.49$~km~s$^{-1}$, a density compression ratio of $r_n=2.93$, a magnetic compression ratio of $r_B=1.62$, an Alfv'enic Mach number of $M_A=4.27$, and a magnetosonic Mach number of $M_{\rm ms}=2.70$. ICME~7 exhibits higher values, with $V_{s,x}=1182.80$~km~s$^{-1}$, $r_n=3.47$, $r_B=2.70$, $M_A=10.42$, and $M_{\rm ms}=4.90$. According to the derived parameters in Table~2, these two events have the highest shock speeds, density and magnetic compression ratios, and Mach numbers among the ten events analyzed. These properties are consistent with favorable conditions for efficient shock-associated particle acceleration. Taken together, the shock parameters, reduced spatial diffusion, enhanced pitch-angle scattering, and coherent magnetic topology suggest that both shock acceleration and magnetic confinement may contribute to the observed energetic-ion enhancements and hard spectra. The fitted spectral indices range from $m=-0.82$ to $-0.64$ for ICME~2 and approximately $m=-0.58$ to $-0.23$ for ICME~7 across the detector configurations (Figure A2). These relatively hard spectra are consistent with efficient energization and transport within the organized magnetic environment; however, the individual contributions of shock acceleration and magnetic confinement cannot be isolated from the present diagnostics alone.

\subsection{Dispersed Upstream Energetic Ion Events}

Among the ten analyzed ICME-driven shock events, seven events (ICMEs 1, 3, 4, 6, 8, 9, and 10) exhibit clear energy-dependent dispersed upstream energetic ion flux variations. In these events, the onset of the upstream ion enhancement is strongly energy-dependent, with higher-energy ions arriving systematically earlier than lower-energy ions. Figure 6 presents a representative example of a dispersed event observed during 2026 January 09--10 (ICME 10). Similar velocity dispersion profiles are consistently observed across all seven dispersed events.

In Figure 6, panels (a)-(d) show the energy-time spectrograms of energetic ion fluxes measured by the PS, IM, EP, and NP detectors of ASPEX-STEPS onboard Aditya-L1. The shaded interval denotes the upstream energetic ion enhancement region preceding the ICME-driven shock arrival, while the vertical dashed lines mark the onset and end of the enhancement interval, with the second dashed line corresponding to the shock arrival at L1. Unlike the dispersion-less events, these seven events display a clear energy-dependent arrival pattern, in which the flux enhancement occurs progressively later with decreasing particle energy. Such behavior reflects velocity dispersion during particle transport along open or weakly organized magnetic field lines from the shock acceleration region toward the spacecraft. The pitch-angle variations superimposed on the spectrograms (black dashed curves) show that energetic ions are detected over a continuously evolving range of pitch angles throughout the upstream interval.

\begin{figure*}[ht!]
\plotone{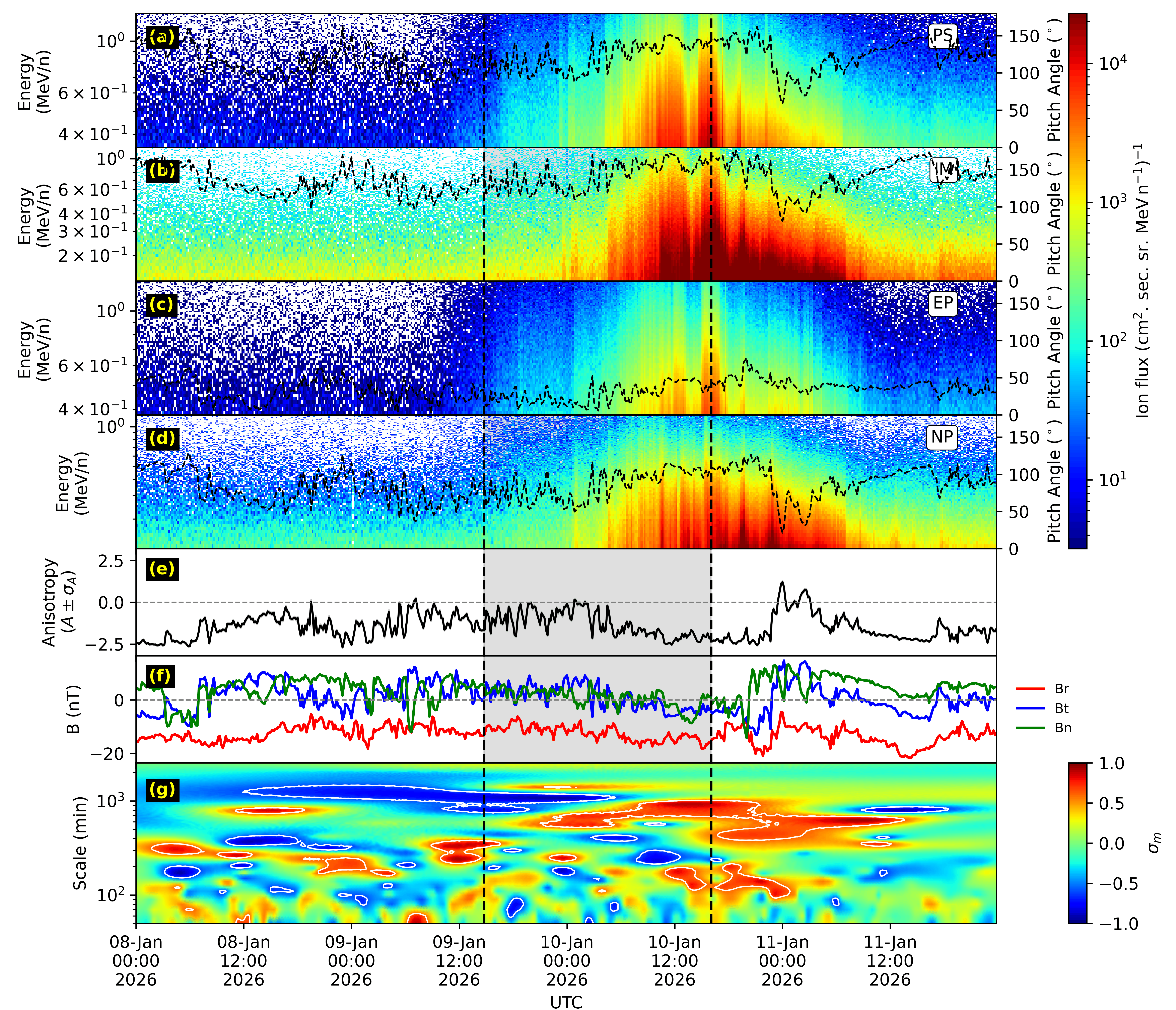}
\caption{Representative example of a dispersed upstream energetic ion enhancement event associated with an ICME-driven shock observed during 2026 January 09--10 (ICME 10). Panels (a)--(d) show energy-time spectrograms of energetic ion fluxes measured by the PS, IM, EP, and NP detectors of ASPEX-STEPS onboard Aditya-L1, respectively, with overplotted dashed black curves representing the corresponding ion pitch angles derived from Aditya-L1 MAG observations. The shaded region marks the upstream energetic ion enhancement interval (from 14:45 UTC on 9 January 2026 to 16:03 UTC on 10 January 2026) ahead of the shock, with the first vertical dashed line marking the enhancement onset and the second vertical dashed line indicating the shock arrival time at L1. Panel (e) presents the first-order anisotropy index ($A$), panel (f) shows the RTN magnetic field components measured by the MAG instrument, and panel (g) displays the scale and time-dependent normalized reduced magnetic helicity ($\sigma_m$) derived using complex Morlet wavelet analysis, where white contours enclose regions with coherent helical magnetic structures ($|\sigma_m| \gtrsim 0.6$).}
\label{fig:general} 
\end{figure*}

The first-order anisotropy index ($A$), shown in panel (e), remains predominantly negative throughout most of the upstream interval, indicating preferential streaming of energetic ions opposite to the local magnetic field direction \citep{Sanderson1985, Tan2012, Dresing2014}. Unlike the dispersion-less events, the anisotropy does not exhibit persistent sign reversals, indicating the absence of prolonged bidirectional trapping or confinement within an upstream flux-rope structure \citep{Malandraki2002, Dresing2016, Ding2025}. Instead, the steady anisotropy pattern suggests that energetic ions are streaming relatively freely along open interplanetary magnetic field lines from the shock front toward the spacecraft \citep{Sanderson1985, Reames1999, Dresing2014}.

The magnetic-field organization of these events was further examined using minimum variance analysis (MVA) and reduced magnetic helicity ($\sigma_m$). The eigenvalue ratios $\lambda_2/\lambda_3$ are 1.60 for ICME~1, 1.85 for ICME~3, 2.28 for ICME~4, 3.44 for ICME~6, 1.53 for ICME~8, 1.83 for ICME~9, and 1.74 for ICME~10. Six of the seven events therefore have $\lambda_2/\lambda_3<3$, indicating comparatively weak separation between the intermediate- and minimum-variance eigenvalues. ICME~6 is an exception, with a ratio of 3.44, demonstrating that a relatively well-constrained variance direction is not exclusive to dispersion-less events. Panel (g) shows that reduced magnetic helicity is generally more intermittent in these dispersed events than in the two dispersion-less events, which exhibit extended intervals with $|\sigma_m|\gtrsim0.6$ over timescales exceeding $10^3$~min. The combined MVA and helicity results suggest that sustained magnetic organization is less evident in most of the dispersed events, although the degree of organization varies from event to event. Consistent with these MVA and helicity diagnostics, 2D Grad--Shafranov (GS) reconstructions (Section \ref{sec:gs_reconstruction}) failed to yield coherent, closed-contour flux-rope topologies for these events, which may be due to fragmented or open magnetic structures that allow particles to escape and disperse freely.

The quasi-linear transport parameters provide additional insight into the differences between the dispersed and dispersion-less events (Table~3). At 150~keV, the parallel mean free paths ($\lambda_\parallel$) for the dispersed events range from 0.0118~AU (ICME~8) to 0.0555~AU (ICME~1), corresponding to spatial diffusion coefficients ($\kappa_\parallel$) of $3.15\times10^{19}$--$1.48\times10^{20}$~cm$^2$~s$^{-1}$ and pitch-angle diffusion coefficients ($\langle D_{\mu\mu}\rangle$) of $4.77\times10^{-4}$--$2.82\times10^{-3}$~s$^{-1}$. At 1.2~MeV, $\lambda_\parallel$ ranges from 0.0128~AU (ICME~8) to 0.0894~AU (ICME~1), $\kappa_\parallel$ spans $9.70\times10^{19}$--$6.76\times10^{20}$~cm$^2$~s$^{-1}$, and $\langle D_{\mu\mu}\rangle$ ranges from $8.38\times10^{-4}$ to $5.14\times10^{-3}$~s$^{-1}$. At both energies, these seven events have larger parallel mean free paths and spatial diffusion coefficients, together with lower pitch-angle diffusion coefficients, than ICME~2 and ICME~7. Within the adopted transport model, this combination is consistent with weaker effective pitch-angle scattering and less restricted transport along the magnetic field. Such conditions can allow energy-dependent propagation to remain evident in the observed onset profiles, whereas stronger scattering and reduced parallel transport may contribute to the nearly simultaneous onsets in the dispersion-less events.

The shock parameters provide further context for the observed differences (Table~2). The radial shock speeds of the dispersed events range from $V_{s,x}=497.49$~km~s$^{-1}$ (ICME~3) to $908.08$~km~s$^{-1}$ (ICME~10). The density compression ratios range from 1.00 to 1.63, the magnetic compression ratios from 1.01 to 1.79, the Alfv\'enic Mach numbers from 1.02 to 3.02, and the fast-magnetosonic Mach numbers from 1.00 to 2.35. Overall, these events exhibit lower compression ratios and Mach numbers than the two dispersion-less events, although the shock speed of ICME~10 slightly exceeds that of ICME~2. The fitted spectral indices (Table~1) are generally more negative than those of the dispersion-less events, ranging from $-1.12$ to $-1.72$ for PS, $-0.94$ to $-1.66$ for IM, $-1.08$ to $-2.38$ for EP, and $-0.86$ to $-1.69$ for NP. For example, ICME~3 has $m=-2.38\pm0.07$ in EP, while ICME~1 has $m=-1.72\pm0.07$ in PS. As the fitted spectrum follows $J(E)\propto E^m$, these more negative indices indicate softer spectra. Taken together, the generally lower shock compression ratios and Mach numbers, weaker effective pitch-angle scattering, and lack of comparably well-defined closed 2D GS contours are consistent with less effective particle confinement and more pronounced energy-dependent propagation, providing a plausible explanation for the clear velocity-dispersion signatures observed across these seven events.

\subsection{Mixed Dispersed and Dispersion-less Upstream Energetic Ion Event}

Among the ten ICME-driven shock events analyzed in this study, one event (ICME~5, observed between 12:00~UTC on 2024 October 09 and 14:49~UTC on 2024 October 10) exhibits a mixed-type upstream energetic-ion enhancement. This event combines characteristics of both dispersed and dispersion-less flux variations, as shown in Figure 7. The shaded region marks the upstream energetic-ion enhancement interval preceding the shock arrival, while the vertical dashed lines indicate the onset and end of the interval, with the second dashed line marking the shock arrival at L1.

Panels (a)--(d) show the energy-time spectrograms measured by the PS, IM, EP, and NP detectors of ASPEX-STEPS onboard Aditya-L1, together with the corresponding particle pitch angles (black dashed curves). During the initial phase, the flux enhancements exhibit a clear energy dependence, with higher-energy ions intensifying earlier than lower-energy ions. As the shock approaches, however, the fluxes across multiple energy channels undergo a rapid enhancement with substantially reduced energy-dependent timing differences. This evolution from dispersed to nearly simultaneous flux enhancements motivates the classification of ICME~5 as a mixed event and distinguishes it from the predominantly dispersed and dispersion-less events discussed above.

\begin{figure*}[ht!]
\plotone{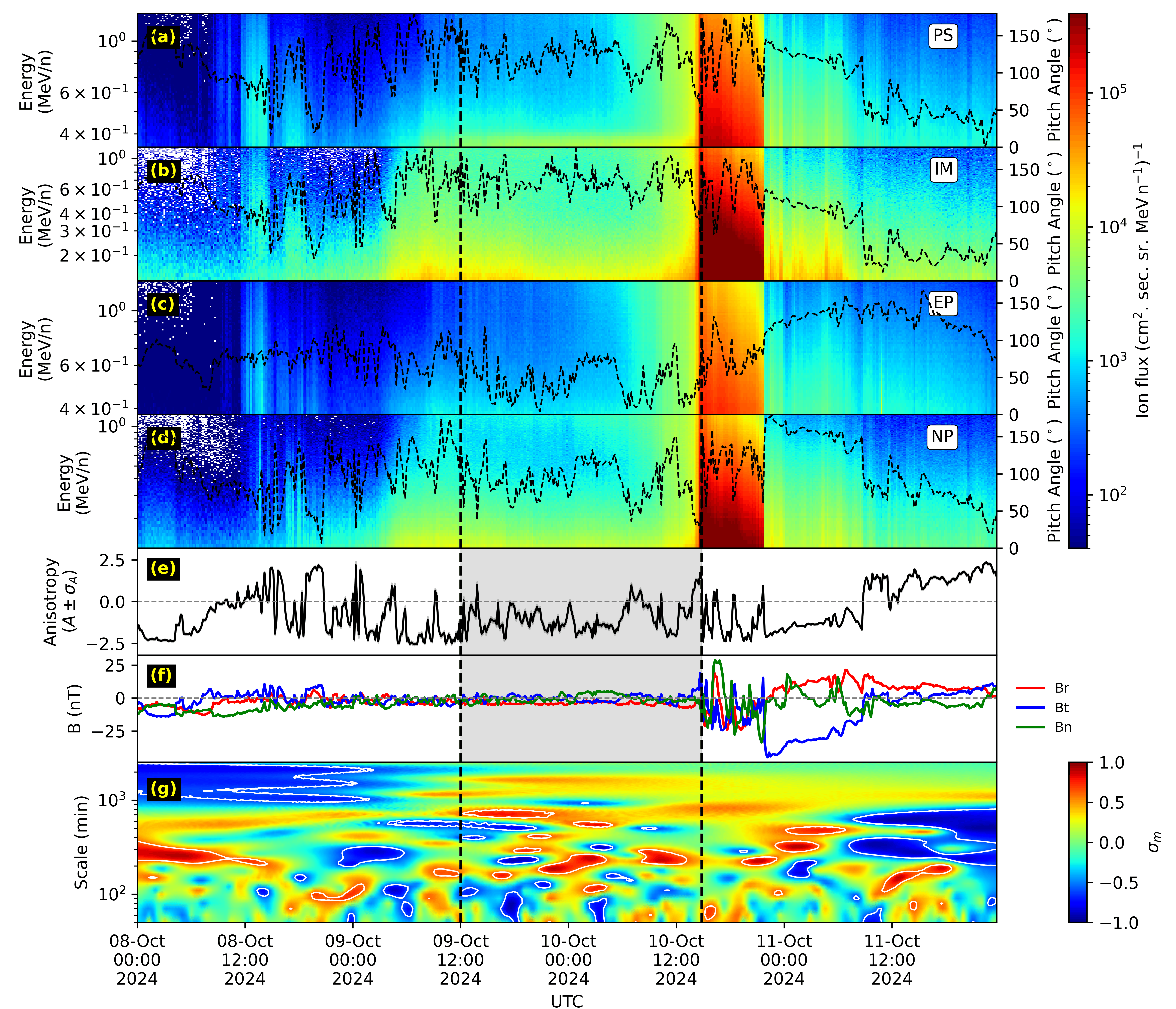}
\caption{This figure represents mixed upstream energetic ion enhancement event associated with an ICME-driven shock observed during 2024 October 09--10 (ICME 5). Panels (a)--(d) show energy-time spectrograms of energetic ion fluxes measured by the PS, IM, EP, and NP detectors of ASPEX-STEPS onboard Aditya-L1, respectively, with overplotted dashed black curves representing the corresponding ion pitch angles derived from Aditya-L1 MAG observations. The shaded region marks the upstream energetic ion enhancement interval (from 12:00 UTC on 9 October 2024 to 14:49 UTC on 10 October 2024) ahead of the shock, with the first vertical dashed line marking the enhancement onset and the second vertical dashed line indicating the shock arrival time at L1. Panel (e) presents the temporal variation of the first-order anisotropy index ($A$), panel (f) shows the RTN magnetic field components measured by the MAG instrument, and panel (g) displays the scale and time-dependent normalized reduced magnetic helicity ($\sigma_m$) derived using complex Morlet wavelet analysis, where regions enclosed by white contours indicate coherent helical magnetic structures ($|\sigma_m| \gtrsim 0.6$).}
\label{fig:general}
\end{figure*}

The temporal evolution of the first-order anisotropy index ($A$), shown in panel (e), provides additional information about the particle transport. During the early phase, the anisotropy remains predominantly negative, indicating preferential streaming opposite to the local magnetic-field direction under the adopted sign convention \citep{Sanderson1985,Tan2012,Dresing2014}. Closer to the shock arrival, the anisotropy exhibits stronger fluctuations and intermittent sign reversals around zero. These variations suggest changing streaming directions and may be consistent with increasingly complex particle transport or localized confinement near the shock \citep{Malandraki2002,Dresing2016,Ding2025}. However, anisotropy variations alone cannot establish magnetic trapping or repeated particle reflections between the shock and surrounding magnetic irregularities.

The magnetic structure was further examined using minimum variance analysis (MVA) and reduced magnetic helicity ($\sigma_m$). MVA yields an eigenvalue ratio of $\lambda_2/\lambda_3=2.61$ for ICME~5, indicating moderate separation between the intermediate and minimum-variance eigenvalues. This value is below the commonly adopted practical guideline of $\lambda_2/\lambda_3\gtrsim3$ for a comparatively well-constrained minimum-variance direction, although the ratio alone cannot determine whether a flux rope is present. Panel (g) shows localized intervals of enhanced reduced magnetic helicity, with $|\sigma_m|\gtrsim0.6$, interspersed with more intermittent and fragmented signatures. Consistent with these diagnostics, the 2D Grad--Shafranov (GS) reconstruction (Section~\ref{sec:gs_reconstruction}) did not yield a comparably well-defined, closed magnetic-flux structure for ICME~5 to those obtained for ICME~2 and ICME~7. Together, these observations suggest an intermediate magnetic environment with localized organization but without a clearly reconstructed, coherent closed-flux-rope cross-section.

The quasi-linear transport parameters provide further insight into this mixed behavior (Table~3). At 150~keV, the parallel mean free path, spatial diffusion coefficient, and pitch-angle diffusion coefficient are $\lambda_\parallel=0.0049$~AU, $\kappa_\parallel=1.30\times10^{19}$~cm$^2$~s$^{-1}$, and $\langle D_{\mu\mu}\rangle=4.85\times10^{-3}$~s$^{-1}$, respectively. At 1.2~MeV, the corresponding values are $\lambda_\parallel=0.0064$~AU, $\kappa_\parallel=4.85\times10^{19}$~cm$^2$~s$^{-1}$, and $\langle D_{\mu\mu}\rangle=9.99\times10^{-3}$~s$^{-1}$. These values fall between the corresponding values for the two dispersion-less events and the ranges obtained for the dispersed events. The relatively short mean free paths and elevated pitch-angle diffusion coefficients, compared with the dispersed events, are consistent with stronger effective scattering, while the transport parameters remain comparable to those of the dispersion-less events. This intermediate transport behavior may contribute to the observed transition from energy-dependent onsets during the early phase to nearly simultaneous enhancements closer to the shock. 

The shock parameters provide additional context for the late-stage enhancement (Table~2). ICME~5 has a radial shock speed of $V_{s,x}=934.93$~km~s$^{-1}$, a density compression ratio of $r_n=3.89$, a magnetic compression ratio of $r_B=1.62$, an Alfv\'enic Mach number of $M_A=2.09$, and a fast-magnetosonic Mach number of $M_{\rm ms}=1.68$. The density compression ratio is the highest among the ten events, whereas the Mach numbers are lower than those of the two dispersion-less events. The high density compression ratio indicates substantial plasma compression across the shock, which may promote shock-associated particle acceleration and contribute to the energization of upstream energetic ions. The fitted spectral indices (Table~1) are $m=-0.71\pm0.16$ (PS), $m=-0.81\pm0.09$ (IM), $m=-0.26\pm0.03$ (EP), and $m=-0.55\pm0.10$ (NP), indicating relatively hard spectra under the convention $J(E)\propto E^m$. The combination of substantial shock compression, intermediate transport properties, evolving anisotropy, and localized magnetic organization is consistent with the mixed character of ICME~5. In particular, the observed transition may reflect the combined influence of shock-associated energization and changing particle transport near the shock, although their individual contributions cannot be isolated from the present diagnostics alone.

\section{Summary and Conclusion}
\label{sec:Summary and Conclusion}

In this study, we investigated the upstream energetic-ion environment associated with ten ICME-driven shocks observed near 1 AU using multidirectional measurements from the ASPEX-STEPS instrument onboard Aditya-L1. Based on their temporal intensity profiles and energy-dependent onset times, the events were classified into three categories: dispersion-less, dispersed, and mixed. To investigate the physical conditions associated with these distinct transport signatures, we combined energetic-ion flux measurements and pitch-angle anisotropy analysis with minimum variance analysis (MVA), normalized reduced magnetic helicity ($\sigma_m$), two-dimensional (2D) Grad-Shafranov (GS) magnetic-field reconstruction, quasi-linear particle-transport diagnostics, and local shock parameters. The results reveal systematic differences in magnetic structure, particle transport, shock properties, and spectral hardness among the three event classes.
\begin{enumerate}
\item \textbf{Dispersionless events:} These two events exhibit nearly simultaneous energetic-ion enhancements across multiple energy channels, extended intervals of enhanced reduced magnetic helicity ($|\sigma_m| \gtrsim 0.6$), and relatively large MVA eigenvalue ratios ($\lambda_2/\lambda_3 = 3.42$ and $3.55$, respectively). Their 2D GS reconstructions yield well-defined closed magnetic-flux contours and distinct core-field enhancements, supporting the presence of coherent, flux-rope-like magnetic structures in the analyzed intervals. The parallel mean free paths are comparatively short, ranging from approximately $0.0037$ to $0.0099$ AU across the two representative energies and events, while the parallel diffusion coefficients are among the lowest and the pitch-angle diffusion coefficients among the highest in the ten events analyzed. Within the adopted transport model, this combination is consistent with stronger effective pitch-angle scattering and more restricted parallel transport. This interpretation is also supported by the directional pitch-angle anisotropy measurements, which show persistently low first-order anisotropies and nearly isotropic or bidirectional energetic-ion distributions upstream of the shocks. The enhanced magnetic fluctuation amplitudes ($\delta B_{\rm rms}/B_0 \simeq 0.58$--$0.66$) further indicate strongly disturbed upstream magnetic conditions. The combination of enhanced turbulence and coherent magnetic topology may therefore contribute to the suppression of field-aligned streaming and the confinement of energetic ions. In particular, the closed magnetic-flux contours and persistent helicity signatures suggest that the coherent magnetic structures may restrict particle escape across magnetic-flux surfaces, in addition to the effects of parallel pitch-angle scattering.

Both events are associated with relatively strong shocks, with radial shock speeds of $903$ and $1183$ km s$^{-1}$, respectively, and elevated compression ratios and Mach numbers. Their energetic-ion spectra are comparatively hard, with spectral indices ranging from $-0.82$ to $-0.64$ for ICME 2 and approximately $-0.81$ to $-0.26$ for ICME 7. The repeated anisotropy sign reversals observed during ICME 7 are consistent with changes in the preferred streaming direction and potentially complex particle transport. Taken together, these signatures suggest that coherent magnetic topology, enhanced pitch-angle scattering, and shock-associated energization may collectively contribute to the observed dispersion-less profiles and hard spectra. However, the individual contributions of particle injection, shock acceleration, and magnetic confinement cannot be uniquely determined from the present observations.

\item \textbf{Dispersed events:} These seven events exhibit clear energy-dependent onset times, with higher-energy ions generally arriving earlier than lower-energy ions. Their reduced magnetic helicity signatures are typically more intermittent than those of the dispersion-less events, and six of the seven events have MVA eigenvalue ratios below the practical guideline of $\lambda_2/\lambda_3 \sim 3$. Their 2D GS reconstructions do not yield magnetic-flux contours comparable to those obtained for dispersion-less events. The transport analysis generally yields larger parallel mean free paths and diffusion coefficients, together with lower pitch-angle diffusion coefficients, than those obtained for the dispersion-less events. The inferred parallel mean free paths span approximately $0.012$--$0.089$ AU across the two representative energies. These estimates, together with the generally lower magnetic fluctuation amplitudes ($\delta B_{\rm rms}/B_0 \simeq 0.11$--$0.52$; median $\sim 0.27$), are consistent with weaker effective pitch-angle scattering and less restricted parallel propagation. In contrast to the nearly isotropic or bidirectional distributions observed during the dispersion-less events, the dispersed events exhibit more pronounced field-aligned anisotropies, consistent with comparatively less restricted particle streaming along the magnetic field. The absence of comparably coherent magnetic structures in the analyzed intervals may also allow more effective particle escape, although perpendicular transport is not independently quantified in the present analysis. The dispersed events generally exhibit lower density and magnetic compression ratios and lower Alfv\'enic and magnetosonic Mach numbers than the dispersion-less events, although individual shock parameters vary across the events. Their spectra are comparatively softer, with spectral indices ranging approximately from $-0.86$ to $-2.38$ across the detector configurations. Collectively, these observations suggest that less coherent magnetic signatures, weaker effective scattering, and energy-dependent propagation are characteristic of the dispersed events.

\item \textbf{Mixed event:} ICME 5 exhibits an evolving transport signature, with initially dispersed energetic-ion arrivals followed by nearly simultaneous enhancements across multiple energy channels as the shock approaches. Its MVA eigenvalue ratio is $\lambda_2/\lambda_3 = 2.61$, and its reduced magnetic helicity exhibits localized intervals of enhanced magnitude amid more intermittent signatures. The 2D GS reconstruction does not yield closed magnetic-flux contours comparable to those obtained for ICMEs 2 and 7, suggesting that the analyzed interval lacks similarly well-defined large-scale magnetic coherence in the reconstruction. Its normalized magnetic fluctuation amplitude ($\delta B_{\rm rms}/B_0 = 0.552$) is nevertheless comparable to those of the dispersion-less events. The event has the largest density compression ratio in the sample ($r_n = 3.89$), although its Alfv\'enic and magnetosonic Mach numbers are lower than those of the dispersion-less events. Its relatively short parallel mean free paths ($\lambda_\parallel = 0.0049$ AU at 150 keV and $0.0064$ AU at 1.2 MeV), together with elevated pitch-angle diffusion coefficients, indicate substantial effective scattering within the adopted transport model. The directional anisotropy measurements also indicate relatively low first-order anisotropies and nearly isotropic or bidirectional ion distributions during the upstream enhancement. These observations suggest that enhanced pitch-angle scattering may suppress field-aligned streaming and favor partial or temporary particle confinement, even though the magnetic reconstruction does not establish a coherent, closed flux-rope structure comparable to those identified in ICMEs 2 and 7. Such conditions may favor partial or temporary particle confinement in the upstream region while allowing continued escape, providing a possible explanation for the coexistence of initially dispersed propagation and subsequent nearly simultaneous flux enhancements. The high density compression ratio may also favor shock-associated particle energization, although compression alone does not establish acceleration efficiency. The relatively hard spectral indices, ranging from approximately $-0.81$ to $-0.26$, may therefore reflect the combined effects of shock-associated energization and energy-dependent transport, with the observed spectrum shaped by the balance between particle retention and escape.

\end{enumerate}

A comparison with the quiet-time suprathermal-ion population previously measured by ASPEX-STEPS provides further context for these results. During quiet periods, the suprathermal-ion spectra were nearly isotropic across the detector viewing directions and exhibited a mean spectral index of $-1.99 \pm 0.06$ \citep{Gupta2025}. In contrast, the upstream energetic-ion spectra associated with ICME-driven shocks exhibit substantial event-to-event variability. The dispersed events span a broad range of spectral slopes, including values comparable to or steeper than the quiet-time spectrum, whereas the dispersion-less and mixed events generally exhibit harder spectra. This spectral hardening, together with the observed anisotropy, magnetic-field structure, and differences in transport parameters, indicates that the upstream energetic-ion population during ICME-driven shocks differs substantially from the quiet-time suprathermal population. The observations are consistent with varying contributions from shock-associated energization and energy-dependent transport in the local magnetic environment.

Overall, this study demonstrates that upstream energetic-ion transport ahead of ICME-driven shocks is associated with systematic variations in magnetic topology, effective pitch-angle scattering, shock properties, and spectral hardness. The dispersion-less events are distinguished by coherent magnetic signatures, well-defined closed contours in 2D GS reconstructions, relatively short parallel mean free paths, and hard spectra. The dispersed events generally exhibit more intermittent magnetic signatures, larger parallel transport coefficients, and pronounced velocity dispersion, while ICME 5 displays an evolving combination of transport characteristics.

These observations support a picture in which energetic-ion transport is influenced by the interplay between parallel pitch-angle scattering, perpendicular escape, and magnetic topology, as also discussed by \citet{Zhao2019}. In particular, enhanced upstream turbulence can suppress field-aligned streaming through more frequent pitch-angle scattering, whereas coherent magnetic structures may additionally restrict particle escape across magnetic-flux surfaces. However, the relative contributions of these processes cannot be separated quantitatively using the present observations and the adopted QLT framework. The QLT-derived transport coefficients should therefore be regarded as first-order, model-dependent estimates, particularly in the strongly turbulent intervals where the weak-turbulence assumption is not uniformly satisfied. To resolve this aspect future studies incorporating nonlinear guiding-center transport and time-dependent focused-transport modeling \citep[e.g.,][]{Zhao2018,Subashchandar2025} will be valuable for refining the absolute diffusion coefficients and investigating the relative roles of parallel and perpendicular transport.

These findings provide a quantitative observational framework for distinguishing energetic-ion transport regimes using multidirectional measurements from Aditya-L1 and highlight the importance of considering shock conditions, magnetic topology, and particle transport together when interpreting energetic-ion enhancements upstream of ICME-driven shocks near 1 AU.

\begin{acknowledgments}
The authors sincerely acknowledge the support provided by various ISRO centres in facilitating the testing, calibration, and technical development of the ASPEX payload. The authors also express their gratitude to the ISRO project, mission, and review teams for their continuous guidance and support. The authors thank the Aditya-L1 Magnetometer (MAG) team for providing the magnetic field data used in this study. The authors also acknowledge the teams responsible for the Wind spacecraft's Magnetic Field Investigation (MFI) and Solar Wind Experiment (SWE) instruments for providing the magnetic field and solar wind plasma data used in this work. We express our sincere gratitude to the Physical Research Laboratory (PRL), Ahmedabad, and the Department of Space (DoS), Government of India, for supporting this research. Y. acknowledges support from the College of Liberal Arts and Sciences at the University of Iowa.
\end{acknowledgments}

\section*{DATA AVAILABILITY STATEMENTS}

The datasets utilized in this study are publicly available. The ASPEX-STEPS measurements from the Aditya-L1 mission are accessible through two repositories: data covering the period from 7 January to 10 May 2024 are available via Zenodo (\doi{10.5281/zenodo.15833452}) \citep{chakrabarty_2025_15833452}, while data acquired from 11 May 2024 onward are distributed through the ISSDC PRADAN portal (\url{https://pradan.issdc.gov.in/al1/}). The Aditya-L1 Magnetometer (MAG) data used in this analysis are also available through the ISSDC PRADAN portal. The Wind spacecraft magnetic-field data from the Magnetic Field Investigation (MFI) and solar wind plasma data from the Solar Wind Experiment (SWE) are publicly available through NASA's Space Physics Data Facility (SPDF), including the Coordinated Data Analysis Web (CDAWeb) interface (\url{https://cdaweb.gsfc.nasa.gov/}).

\appendix
\setcounter{figure}{0}
\renewcommand{\thefigure}{A\arabic{figure}}

\section{Energy Spectra of Dispersed Events}

\begin{figure*}[ht!]
\centering
\includegraphics[width=\textwidth]{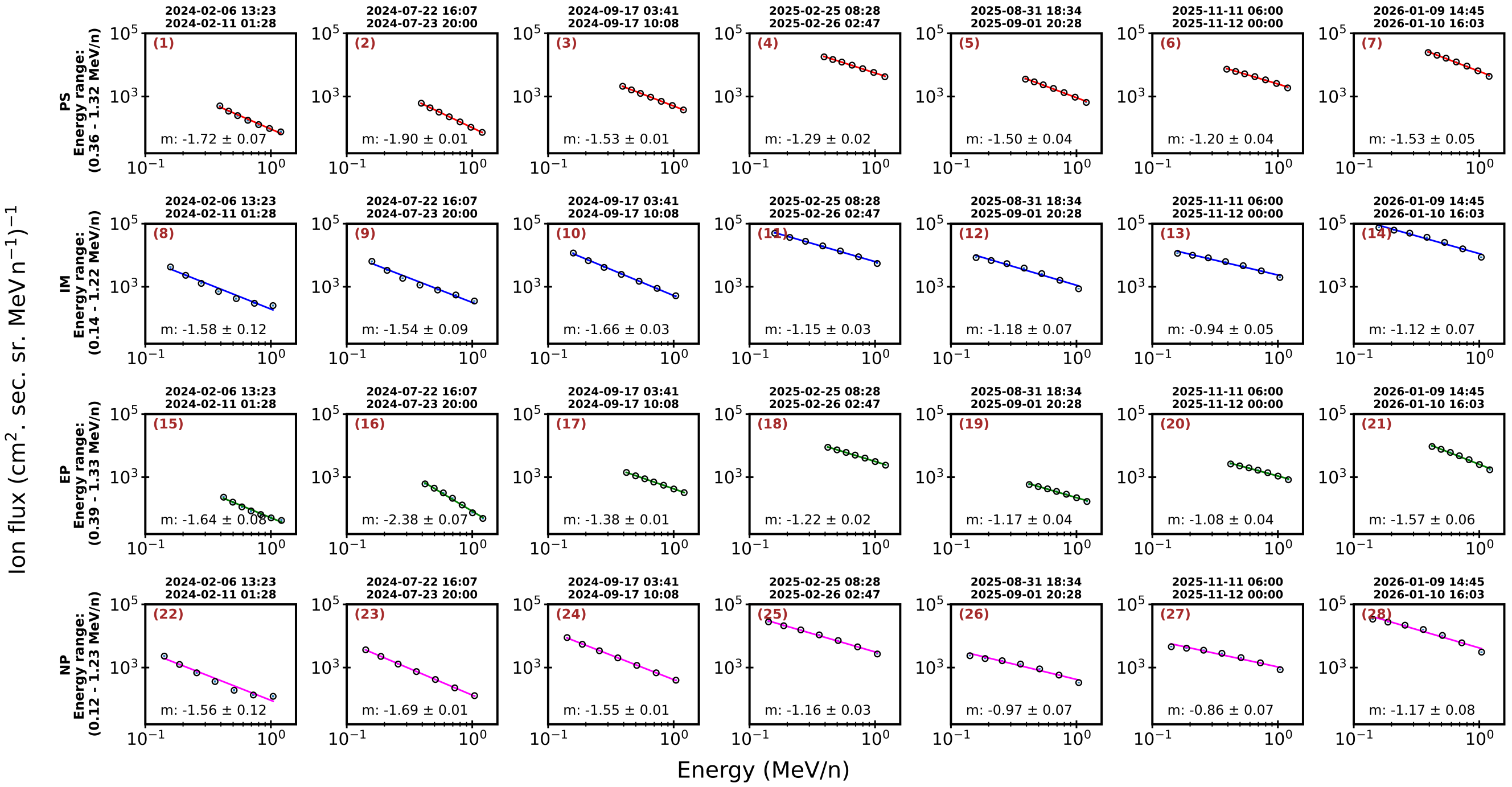}
\caption{Energy spectra of upstream accelerated ions for the seven dispersed ICME events across different ASPEX-STEPS detector configurations. Columns (1) to (7) correspond to individual analyzed events, ordered chronologically. Rows from top to bottom display data from the PS (red), IM (blue), EP (green), and NP (magenta) detectors, respectively. In each panel, open circles denote measured ion intensities with associated $1\sigma$ observational error bars, while the solid lines represent the best-fit power-law profiles. The calculated spectral index ($m$) and its corresponding fitting uncertainty are explicitly labeled at the bottom of each panel.}
\label{fig:appendix_spectra}
\end{figure*}

Figure~\ref{fig:appendix_spectra} presents the power-law fits to the upstream energetic-ion spectra for the seven dispersed ICME events, obtained using the PS, IM, EP, and NP detector configurations of ASPEX-STEPS onboard Aditya-L1. Following the methodology described in Section~3.2, the spectral index ($m$) was determined by fitting a linear relation to the logarithmic differential flux as a function of energy over the selected energy interval for each detector configuration. The derived spectral indices range from $-2.38 \pm 0.07$ to $-0.86 \pm 0.07$, indicating substantial variability in spectral hardness among events and viewing directions. These variations may reflect differences in the local shock conditions and energetic-ion transport, including energy-dependent propagation and pitch-angle scattering.

\section{Energy Spectra of Dispersionless and Mixed Events}

\begin{figure*}[ht!]
\centering
\includegraphics[width=\textwidth]{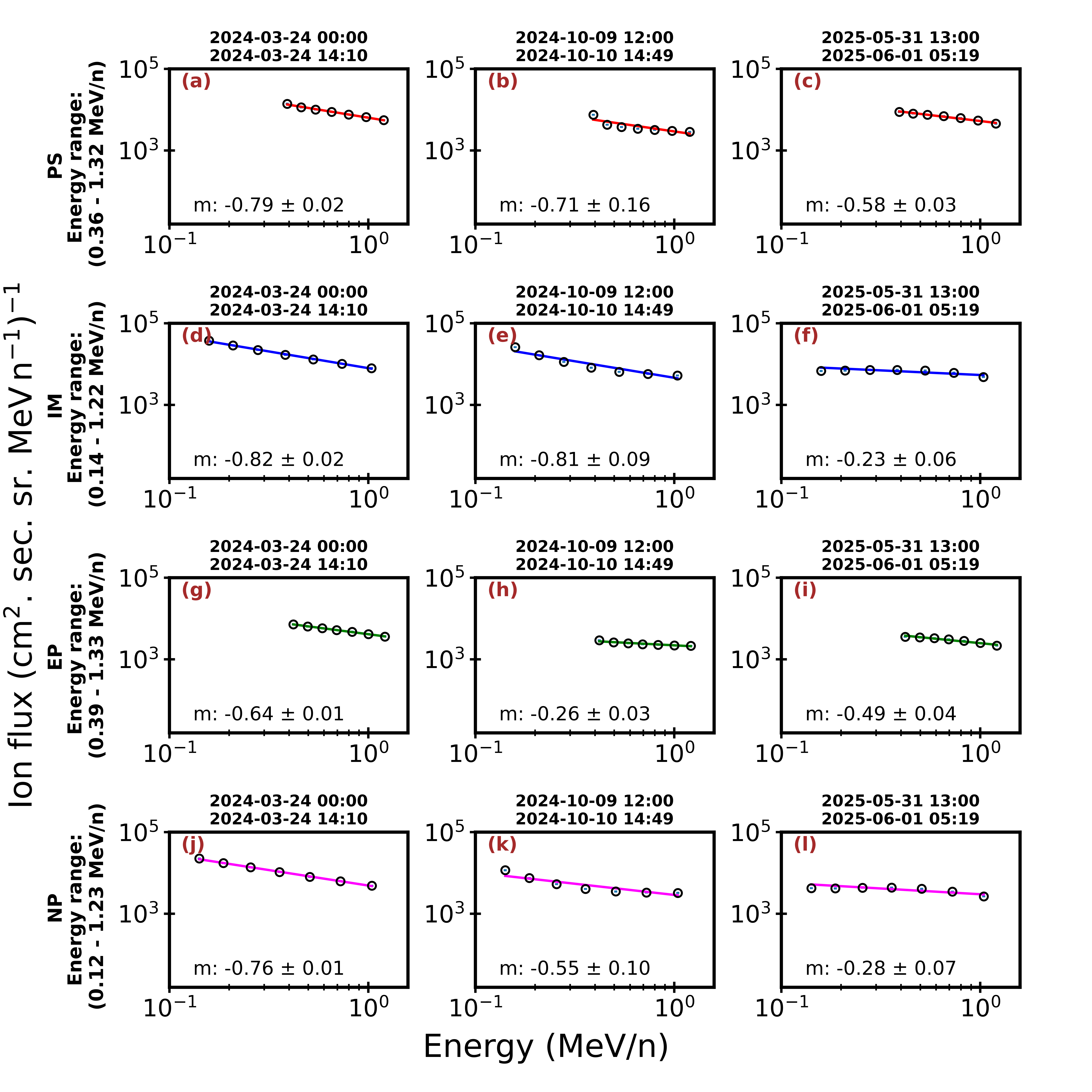}
\caption{Power-law spectral fits of upstream energetic ion fluxes for the dispersion-less and mixed ICME events. The first and third columns correspond to the dispersion-less events, while the middle column represents the mixed event. Rows from top to bottom shows observations from the PS (red), IM (blue), EP (green), and NP (magenta) detectors of ASPEX-STEPS onboard \textit{Aditya-L1}. Open circles denote the measured ion fluxes with corresponding $1\sigma$ uncertainties, and solid lines represent the best-fit power-law spectra. The derived spectral index ($m$) and fitting uncertainty are indicated in each panel.}
\label{fig:appendix_spectra_dispersionless}
\end{figure*}

Figure~\ref{fig:appendix_spectra_dispersionless} presents the power-law fits to the upstream energetic-ion spectra for the two dispersionless events (Events 2 and 7 in Table~1) and the mixed event (Event 5 in Table~1). Following the methodology described in Section~3.2, the spectral indices ($m$) were determined by fitting a linear relation to the logarithmic differential flux as a function of energy over the selected energy intervals for the PS, IM, EP, and NP detector configurations of ASPEX-STEPS onboard Aditya-L1. The derived spectral indices range from $-0.23 \pm 0.06$ to $-0.82 \pm 0.02$, indicating comparatively hard energetic-ion spectra across these events. As discussed in the main text, these hard spectra occur alongside distinct magnetic-topology signatures, variations in pitch-angle anisotropy, and differences in particle-transport and shock parameters. The combined observations suggest that the spectral characteristics may reflect the interplay between shock-associated energization and energy-dependent particle transport, potentially including partial confinement in structured magnetic environments. The complete spectral fits provide further quantitative support for the differences in energetic-ion populations and transport characteristics discussed in the main text.

\begin{thebibliography}{}
\expandafter\ifx\csname natexlab\endcsname\relax\def\natexlab#1{#1}\fi
\providecommand{\url}[1]{\href{#1}{#1}}
\providecommand{\dodoi}[1]{doi:~\href{http://doi.org/#1}{\nolinkurl{#1}}}
\providecommand{\doeprint}[1]{\href{http://ascl.net/#1}{\nolinkurl{http://ascl.net/#1}}}
\providecommand{\doarXiv}[1]{\href{https://arxiv.org/abs/#1}{\nolinkurl{https://arxiv.org/abs/#1}}}

\bibitem[{B. Abraham-Shrauner(1972)Abraham-Shrauner}]{AbrahamShrauner1972}
Abraham-Shrauner, B. 1972, \bibinfo{title}{Determination of magnetohydrodynamic shock normals,} Journal of Geophysical Research, 77, 736, \dodoi{10.1029/JA077i004p00736}

\bibitem[{B. Abraham-Shrauner \& S.~H. Yun(1976)Abraham-Shrauner \& Yun}]{AbrahamShraunerYun1976}
Abraham-Shrauner, B., \& Yun, S.~H. 1976, \bibinfo{title}{Interplanetary shocks seen by Ames Research Center's Pioneer 6 and 7,} Journal of Geophysical Research, 81, 2097, \dodoi{10.1029/JA081i013p02097}

\bibitem[{M. Acu{\~n}a {et~al.}(1995)Acu{\~n}a, Ogilvie, Baker, Curtis, Fairfield, \& Mish}]{Acuna1995}
Acu{\~n}a, M., Ogilvie, K., Baker, D., {et~al.} 1995, \bibinfo{title}{The global geospace science program and its investigations,} Space Science Reviews, 71, 5, \dodoi{10.1007/BF00751323}

\bibitem[{K.~A. {Anderson} \& W.~M. {Dougherty}(1986){Anderson} \& {Dougherty}}]{Anderson1986}
{Anderson}, K.~A., \& {Dougherty}, W.~M. 1986, \bibinfo{title}{{A Spatially Confined Long-Lived Stream of Solar Particles},} \solphys, 103, 165, \dodoi{10.1007/BF00154866}

\bibitem[{J.~J. Blanco {et~al.}(2013)Blanco, Hidalgo, G{\'o}mez-Herrero, Rodr{\'i}guez-Pacheco, Heber, Wimmer-Schweingruber, \& Mart{\'i}n}]{Blanco2013}
Blanco, J.~J., Hidalgo, M.~A., G{\'o}mez-Herrero, R., {et~al.} 2013, \bibinfo{title}{Energetic-particle-flux decreases related to magnetic cloud passages as observed by the Helios 1 and 2 spacecraft,} Astronomy \& Astrophysics, 556, A146, \dodoi{10.1051/0004-6361/201321739}

\bibitem[{J.~E. {Borovsky}(2008){Borovsky}}]{Borovsky2008}
{Borovsky}, J.~E. 2008, \bibinfo{title}{{Flux tube texture of the solar wind: Strands of the magnetic carpet at 1 AU?},} Journal of Geophysical Research (Space Physics), 113, A08110, \dodoi{10.1029/2007JA012684}

\bibitem[{R. {Bruno} {et~al.}(2001){Bruno}, {Carbone}, {Veltri}, {Pietropaolo}, \& {Bavassano}}]{Bruno2001}
{Bruno}, R., {Carbone}, V., {Veltri}, P., {Pietropaolo}, E., \& {Bavassano}, B. 2001, \bibinfo{title}{{Identifying intermittency events in the solar wind},} \planss, 49, 1201, \dodoi{10.1016/S0032-0633(01)00061-7}

\bibitem[{A. {Buttighoffer}(1998){Buttighoffer}}]{Buttighoffer1998}
{Buttighoffer}, A. 1998, \bibinfo{title}{{Solar electron beams associated with radio type III bursts},} \aap, 335, 295

\bibitem[{H.~V. {Cane} {et~al.}(1986){Cane}, {McGuire}, \& {von Rosenvinge}}]{Cane1986}
{Cane}, H.~V., {McGuire}, R.~E., \& {von Rosenvinge}, T.~T. 1986, \bibinfo{title}{{Two Classes of Solar Energetic Particle Events Associated with Impulsive and Long-Duration Soft X-Ray Flares},} \apj, 301, 448, \dodoi{10.1086/163913}

\bibitem[{M.~L. {Cartwright} \& M.~B. {Moldwin}(2010){Cartwright} \& {Moldwin}}]{Cartwright2010}
{Cartwright}, M.~L., \& {Moldwin}, M.~B. 2010, \bibinfo{title}{{Heliospheric evolution of solar wind small-scale magnetic flux ropes},} Journal of Geophysical Research (Space Physics), 115, A08102, \dodoi{10.1029/2009JA014271}

\bibitem[{D. Chakrabarty(2025)Chakrabarty}]{chakrabarty_2025_15833452}
Chakrabarty, D. 2025, These datasets contain ASPEX-STEPS level1a data onboard Aditya-L1 spacecraft from 7th January to 10th May 2024., Zenodo, \dodoi{10.5281/zenodo.15833452}

\bibitem[{X. Chen {et~al.}(2022)Chen, Giacalone, \& Guo}]{Chen2022}
Chen, X., Giacalone, J., \& Guo, F. 2022, \bibinfo{title}{Solar Energetic Particle Acceleration at a Spherical Shock with the Shock Normal Angle $\theta$ B n Evolving in Space and Time,} The Astrophysical Journal, 941, 23, \dodoi{10.3847/1538-4357/ac9f43}

\bibitem[{E.~E. {Chollet} \& J. {Giacalone}(2008){Chollet} \& {Giacalone}}]{Chollet2008}
{Chollet}, E.~E., \& {Giacalone}, J. 2008, \bibinfo{title}{{Multispacecraft Analysis of Energetic Ion Flux Dropouts},} \apj, 688, 1368, \dodoi{10.1086/592378}

\bibitem[{P. Chuychai {et~al.}(2005)Chuychai, Ruffolo, Matthaeus, \& Rowlands}]{Chuychai2005}
Chuychai, P., Ruffolo, D., Matthaeus, W., \& Rowlands, G. 2005, \bibinfo{title}{Suppressed diffusive escape of topologically trapped magnetic field lines,} The Astrophysical Journal Letters, 633, L49, \dodoi{10.1086/498137}

\bibitem[{E.~W. {Cliver} \& A.~G. {Ling}(2009){Cliver} \& {Ling}}]{Cliver2000}
{Cliver}, E.~W., \& {Ling}, A.~G. 2009, \bibinfo{title}{{Low-Frequency Type III Bursts and Solar Energetic Particle Events},} \apj, 690, 598, \dodoi{10.1088/0004-637X/690/1/598}

\bibitem[{M. {Desai} \& J. {Giacalone}(2016){Desai} \& {Giacalone}}]{Desai2016}
{Desai}, M., \& {Giacalone}, J. 2016, \bibinfo{title}{{Large gradual solar energetic particle events},} Living Reviews in Solar Physics, 13, 3, \dodoi{10.1007/s41116-016-0002-5}

\bibitem[{Z. Ding {et~al.}(2025)Ding, Wimmer-Schweingruber, Chen, Zhao, Kollhoff, K{\"u}hl, Yang, Berger, Heidrich-Meisner, Rodriguez-Pacheco, {et~al.}}]{Ding2025}
Ding, Z., Wimmer-Schweingruber, R.~F., Chen, Y., {et~al.} 2025, \bibinfo{title}{Bidirectional anisotropic solar energetic particle events observed by Solar Orbiter,} Astronomy \& Astrophysics, 701, A123, \dodoi{10.1051/0004-6361/202556098}

\bibitem[{N. {Dresing} {et~al.}(2016){Dresing}, {G{\'o}mez-Herrero}, {Heber}, {Hidalgo}, {Klassen}, {Temmer}, \& {Veronig}}]{Dresing2016}
{Dresing}, N., {G{\'o}mez-Herrero}, R., {Heber}, B., {et~al.} 2016, \bibinfo{title}{{Injection of solar energetic particles into both loop legs of a magnetic cloud},} \aap, 586, A55, \dodoi{10.1051/0004-6361/201527347}

\bibitem[{N. {Dresing} {et~al.}(2014){Dresing}, {G{\'o}mez-Herrero}, {Heber}, {Klassen}, {Malandraki}, {Dr{\"o}ge}, \& {Kartavykh}}]{Dresing2014}
{Dresing}, N., {G{\'o}mez-Herrero}, R., {Heber}, B., {et~al.} 2014, \bibinfo{title}{{Statistical survey of widely spread out solar electron events observed with STEREO and ACE with special attention to anisotropies},} \aap, 567, A27, \dodoi{10.1051/0004-6361/201423789}

\bibitem[{N. {Dresing} {et~al.}(2012){Dresing}, {G{\'o}mez-Herrero}, {Klassen}, {Heber}, {Kartavykh}, \& {Dr{\"o}ge}}]{Dresing2012}
{Dresing}, N., {G{\'o}mez-Herrero}, R., {Klassen}, A., {et~al.} 2012, \bibinfo{title}{{The Large Longitudinal Spread of Solar Energetic Particles During the 17 January 2010 Solar Event},} \solphys, 281, 281, \dodoi{10.1007/s11207-012-0049-y}

\bibitem[{N. {Dresing} {et~al.}(2023){Dresing}, {Rodr{\'\i}guez-Garc{\'\i}a}, {Jebaraj}, {Warmuth}, {Wallace}, {Balmaceda}, {Podladchikova}, {Strauss}, {Kouloumvakos}, {Palmroos}, {Krupar}, {Gieseler}, {Xu}, {Mitchell}, {Cohen}, {de Nolfo}, {Palmerio}, {Carcaboso}, {Kilpua}, {Trotta}, {Auster}, {Asvestari}, {da Silva}, {Dr{\"o}ge}, {Getachew}, {G{\'o}mez-Herrero}, {Grande}, {Heyner}, {Holmstr{\"o}m}, {Huovelin}, {Kartavykh}, {Laurenza}, {Lee}, {Mason}, {Maksimovic}, {Mieth}, {Murakami}, {Oleynik}, {Pinto}, {Pulupa}, {Richter}, {Rodr{\'\i}guez-Pacheco}, {S{\'a}nchez-Cano}, {Schuller}, {Ueno}, {Vainio}, {Vecchio}, {Veronig}, \& {Wijsen}}]{Dresing2023}
{Dresing}, N., {Rodr{\'\i}guez-Garc{\'\i}a}, L., {Jebaraj}, I.~C., {et~al.} 2023, \bibinfo{title}{{The 17 April 2021 widespread solar energetic particle event},} \aap, 674, A105, \dodoi{10.1051/0004-6361/202345938}

\bibitem[{J. Giacalone(2012)Giacalone}]{Giacalone2012}
Giacalone, J. 2012, \bibinfo{title}{Energetic charged particles associated with strong interplanetary shocks,} The Astrophysical Journal, 761, 28, \dodoi{10.1088/0004-637X/761/1/28}

\bibitem[{J. Giacalone(2017)Giacalone}]{Giacalone2017}
Giacalone, J. 2017, \bibinfo{title}{The acceleration of charged particles at a spherical shock moving through an irregular magnetic field,} The Astrophysical Journal, 848, 123, \dodoi{10.3847/1538-4357/aa8df1}

\bibitem[{J. Giacalone \& J. Jokipii(1999)Giacalone \& Jokipii}]{Giacalone1999}
Giacalone, J., \& Jokipii, J. 1999, \bibinfo{title}{The transport of cosmic rays across a turbulent magnetic field,} The astrophysical journal, 520, 204, \dodoi{10.1086/307452}

\bibitem[{J. {Giacalone} {et~al.}(2020){Giacalone}, {Mitchell}, {Allen}, {Hill}, {McNutt}, {Szalay}, {Desai}, {Rouillard}, {Kouloumvakos}, {McComas}, {Christian}, {Schwadron}, {Wiedenbeck}, {Bale}, {Brown}, {Case}, {Chen}, {Cohen}, {Joyce}, {Kasper}, {Klein}, {Korreck}, {Larson}, {Livi}, {Leske}, {MacDowall}, {Matthaeus}, {Mewaldt}, {Nieves-Chinchilla}, {Pulupa}, {Roelof}, {Stevens}, {Szabo}, \& {Whittlesey}}]{Giacalone2020}
{Giacalone}, J., {Mitchell}, D.~G., {Allen}, R.~C., {et~al.} 2020, \bibinfo{title}{{Solar Energetic Particles Produced by a Slow Coronal Mass Ejection at {\ensuremath{\sim}}0.25 au},} \apjs, 246, 29, \dodoi{10.3847/1538-4365/ab5221}

\bibitem[{J.~P. Goedbloed \& S. Poedts(2004)Goedbloed \& Poedts}]{GoedbloedPoedts2004}
Goedbloed, J.~P., \& Poedts, S. 2004, Principles of Magnetohydrodynamics: With Applications to Laboratory and Astrophysical Plasmas (Cambridge: Cambridge University Press), \dodoi{10.1017/CBO9780511616945}

\bibitem[{R. G{\'o}mez-Herrero {et~al.}(2017)G{\'o}mez-Herrero, Dresing, Klassen, Heber, Temmer, Veronig, Bu{\v{c}}{\'\i}k, Hidalgo, Carcaboso, Blanco, {et~al.}}]{GomezHerrero2017}
G{\'o}mez-Herrero, R., Dresing, N., Klassen, A., {et~al.} 2017, \bibinfo{title}{Sunward-propagating solar energetic electrons inside multiple interplanetary flux ropes,} The Astrophysical Journal, 840, 85, \dodoi{10.3847/1538-4357/aa6c5c}

\bibitem[{N. {Gopalswamy} {et~al.}(2008){Gopalswamy}, {Yashiro}, {Akiyama}, {M{\"a}kel{\"a}}, {Xie}, {Kaiser}, {Howard}, \& {Bougeret}}]{Gopalswamy2008}
{Gopalswamy}, N., {Yashiro}, S., {Akiyama}, S., {et~al.} 2008, \bibinfo{title}{{Coronal mass ejections, type II radio bursts, and solar energetic particle events in the SOHO era},} Annales Geophysicae, 26, 3033, \dodoi{10.5194/angeo-26-3033-2008}

\bibitem[{J. Gosling {et~al.}(2004)Gosling, Skoug, McComas, \& Mazur}]{Gosling2004}
Gosling, J., Skoug, R.~M., McComas, D., \& Mazur, J. 2004, \bibinfo{title}{Correlated dispersionless structure in suprathermal electrons and solar energetic ions in the solar wind,} The Astrophysical Journal, 614, 412, \dodoi{10.1086/423368}

\bibitem[{S. Goyal {et~al.}(2018)Goyal, Kumar, Janardhan, Vadawale, Sarkar, Shanmugam, Subramanian, Bapat, Chakrabarty, Adhyaru, {et~al.}}]{Goyal2018}
Goyal, S., Kumar, P., Janardhan, P., {et~al.} 2018, \bibinfo{title}{Aditya solarwind particle experiment (ASPEX) onboard the Aditya-L1 mission,} Planetary and Space Science, 163, 42, \dodoi{10.1016/j.pss.2018.04.008}

\bibitem[{S.~K. Goyal {et~al.}(2025)Goyal, Tiwari, Patel, Shanmugam, Vadawale, Chakrabarty, Sebastian, Dalal, Sharma, Sarkar, {et~al.}}]{Goyal_et_al_2025}
Goyal, S.~K., Tiwari, N.~K., Patel, A.~R., {et~al.} 2025, \bibinfo{title}{Aditya Solar Wind Particle Experiment on Board Aditya--L1: The Supra-Thermal and Energetic Particle Spectrometer,} Solar Physics, 300, 35

\bibitem[{F. Guo {et~al.}(2010)Guo, Jokipii, \& Kota}]{Guo2010}
Guo, F., Jokipii, J.~R., \& Kota, J. 2010, \bibinfo{title}{Particle acceleration by collisionless shocks containing large-scale magnetic-field variations,} The Astrophysical Journal, 725, 128, \dodoi{10.1088/0004-637X/725/1/128}

\bibitem[{A. Gupta {et~al.}(2025)Gupta, Chakrabarty, Vadawale, Sarkar, Dalal, Goyal, Sebastian, Janardhan, Srivastava, Shanmugam, {et~al.}}]{Gupta2025}
Gupta, A., Chakrabarty, D., Vadawale, S., {et~al.} 2025, \bibinfo{title}{Multidirectional Investigations on Quiet Time Suprathermal Ions Measured by ASPEX-STEPS on Board Aditya L1,} The Astrophysical Journal, 995, 92, \dodoi{10.3847/1538-4357/ae1a46}

\bibitem[{L.-N. Hau \& B.~U.~{\"O}. Sonnerup(1999)Hau \& Sonnerup}]{HauSonnerup1999}
Hau, L.-N., \& Sonnerup, B. U.~{\"O}. 1999, \bibinfo{title}{Two-dimensional structures in the solar wind,} Journal of Geophysical Research: Space Physics, 104, 6899

\bibitem[{G.~C. Ho {et~al.}(2022)Ho, Mason, Allen, Wimmer-Schweingruber, Rodr{\'i}guez-Pacheco, \& G{\'o}mez-Herrero}]{Ho2022}
Ho, G.~C., Mason, G.~M., Allen, R.~C., {et~al.} 2022, \bibinfo{title}{Interplanetary Ion Flux Dropouts Across Multiple {$^{3}$He}-Rich Events,} Frontiers in Astronomy and Space Sciences, 9, 939799, \dodoi{10.3389/fspas.2022.939799}

\bibitem[{Q. Hu \& B.~U.~{\"O}. Sonnerup(2001)Hu \& Sonnerup}]{HuSonnerup2001}
Hu, Q., \& Sonnerup, B. U.~{\"O}. 2001, \bibinfo{title}{Reconstruction of magnetic flux ropes in the solar wind,} Geophysical Research Letters, 28, 467, \dodoi{10.1029/2000GL012232}

\bibitem[{Q. Hu \& B.~U.~{\"O}. Sonnerup(2002)Hu \& Sonnerup}]{HuSonnerup2002}
Hu, Q., \& Sonnerup, B. U.~{\"O}. 2002, \bibinfo{title}{Reconstruction of magnetic clouds in the solar wind: Orientations and configurations,} Journal of Geophysical Research: Space Physics, 107, \dodoi{10.1029/2001JA000293}

\bibitem[{K. Huttunen-Heikinmaa {et~al.}(2005)Huttunen-Heikinmaa, Valtonen, \& Laitinen}]{HuttunenHeikinmaa2005}
Huttunen-Heikinmaa, K., Valtonen, E., \& Laitinen, T. 2005, \bibinfo{title}{Proton and helium release times in SEP events observed with SOHO/ERNE,} Astronomy \& Astrophysics, 442, 673, \dodoi{10.1051/0004-6361:20042620}

\bibitem[{P. Janardhan {et~al.}(2017)Janardhan, Vadawale, Bapat, Subramanian, Chakrabarty, Kumar, Sarkar, Srivastava, Thampi, Yadav, {et~al.}}]{Janardhan_et_al_2017}
Janardhan, P., Vadawale, S., Bapat, B., {et~al.} 2017, \bibinfo{title}{Probing the heliosphere using in situ payloads on-board Aditya-L1,} Current Science, 620

\bibitem[{I.~C. {Jebaraj} {et~al.}(2024){Jebaraj}, {Agapitov}, {Krasnoselskikh}, {Vuorinen}, {Gedalin}, {Choi}, {Palmerio}, {Wijsen}, {Dresing}, {Cohen}, {Kouloumvakos}, {Balikhin}, {Vainio}, {Kilpua}, {Afanasiev}, {Verniero}, {Mitchell}, {Trotta}, {Hill}, {Raouafi}, \& {Bale}}]{Jebaraj2024}
{Jebaraj}, I.~C., {Agapitov}, O., {Krasnoselskikh}, V., {et~al.} 2024, \bibinfo{title}{{Acceleration of Electrons and Ions by an ``Almost'' Astrophysical Shock in the Heliosphere},} \apjl, 968, L8, \dodoi{10.3847/2041-8213/ad4daa}

\bibitem[{J. Jokipii(1966)Jokipii}]{Jokipii1966b}
Jokipii, J. 1966, \bibinfo{title}{A model of Fermi acceleration at shock fronts with an application to the earth's bow shock,} Astrophysical Journal, vol. 143, p. 961, 143, 961

\bibitem[{J. Jokipii(1982)Jokipii}]{Jokipii1982}
Jokipii, J. 1982, \bibinfo{title}{Particle drift, diffusion, and acceleration at shocks,} Astrophysical Journal, Part 1, vol. 255, Apr. 15, 1982, p. 716-720., 255, 716

\bibitem[{J.~R. {Jokipii}(1966){Jokipii}}]{Jokipii1966}
{Jokipii}, J.~R. 1966, \bibinfo{title}{{Cosmic-Ray Propagation. I. Charged Particles in a Random Magnetic Field},} \apj, 146, 480, \dodoi{10.1086/148912}

\bibitem[{S.~W. {Kahler}(2007){Kahler}}]{Kahler2007}
{Kahler}, S.~W. 2007, \bibinfo{title}{{Solar Sources of Heliospheric Energetic Electron Events{\textemdash}Shocks or Flares?},} \ssr, 129, 359, \dodoi{10.1007/s11214-007-9143-0}

\bibitem[{S.~W. {Kahler} \& A.~J. {Hundhausen}(1992){Kahler} \& {Hundhausen}}]{Kahler1992}
{Kahler}, S.~W., \& {Hundhausen}, A.~J. 1992, \bibinfo{title}{{The Magnetic Topology of Solar Coronal Structures Following Mass Ejections},} \jgr, 97, 1619, \dodoi{10.1029/91JA02402}

\bibitem[{L.~Y. {Khoo} {et~al.}(2024){Khoo}, {S{\'a}nchez-Cano}, {Lee}, {Rodr{\'\i}guez-Garc{\'\i}a}, {Kouloumvakos}, {Palmerio}, {Carcaboso}, {Lario}, {Dresing}, {Cohen}, {McComas}, {Lynch}, {Fraschetti}, {Jebaraj}, {Mitchell}, {Nieves-Chinchilla}, {Krupar}, {Pacheco}, {Giacalone}, {Auster}, {Benkhoff}, {Bonnin}, {Christian}, {Ehresmann}, {Fedeli}, {Fischer}, {Heyner}, {Holmstr{\"o}m}, {Leske}, {Maksimovic}, {Mieth}, {Oleynik}, {Pinto}, {Richter}, {Rodr{\'\i}guez-Pacheco}, {Schwadron}, {Schmid}, {Telloni}, {Vecchio}, \& {Wiedenbeck}}]{Khoo2024}
{Khoo}, L.~Y., {S{\'a}nchez-Cano}, B., {Lee}, C.~O., {et~al.} 2024, \bibinfo{title}{{Multispacecraft Observations of a Widespread Solar Energetic Particle Event on 2022 February 15─16},} \apj, 963, 107, \dodoi{10.3847/1538-4357/ad167f}

\bibitem[{A.~V. Khrabrov \& B.~U.~{\"O}. Sonnerup(1998)Khrabrov \& Sonnerup}]{KhrabrovSonnerup1998}
Khrabrov, A.~V., \& Sonnerup, B. U.~{\"O}. 1998, \bibinfo{title}{DeHoffmann--Teller analysis,} in Analysis Methods for Multi-Spacecraft Data, ed. G.~Paschmann \& P.~W. Daly (Bern: ISSI), 221--248

\bibitem[{K.-L. {Klein} {et~al.}(2005){Klein}, {Krucker}, {Trottet}, \& {Hoang}}]{Klein2005}
{Klein}, K.-L., {Krucker}, S., {Trottet}, G., \& {Hoang}, S. 2005, \bibinfo{title}{{Coronal phenomena at the release of solar energetic electron events},} \aap, 431, 1047, \dodoi{10.1051/0004-6361:20041258}

\bibitem[{X. Kong {et~al.}(2017)Kong, Guo, Giacalone, Li, \& Chen}]{Kong2017}
Kong, X., Guo, F., Giacalone, J., Li, H., \& Chen, Y. 2017, \bibinfo{title}{The acceleration of high-energy protons at coronal shocks: the effect of large-scale streamer-like magnetic field structures,} The Astrophysical Journal, 851, 38, \dodoi{10.3847/1538-4357/aa97d7}

\bibitem[{A. {Kouloumvakos} {et~al.}(2019){Kouloumvakos}, {Rouillard}, {Wu}, {Vainio}, {Vourlidas}, {Plotnikov}, {Afanasiev}, \& {{\"O}nel}}]{Kouloumvakos2019}
{Kouloumvakos}, A., {Rouillard}, A.~P., {Wu}, Y., {et~al.} 2019, \bibinfo{title}{{Connecting the Properties of Coronal Shock Waves with Those of Solar Energetic Particles},} \apj, 876, 80, \dodoi{10.3847/1538-4357/ab15d7}

\bibitem[{A. {Kouloumvakos} {et~al.}(2022){Kouloumvakos}, {Kwon}, {Rodr{\'\i}guez-Garc{\'\i}a}, {Lario}, {Dresing}, {Kilpua}, {Vainio}, {T{\"o}r{\"o}k}, {Plotnikov}, {Rouillard}, {Downs}, {Linker}, {Malandraki}, {Pinto}, {Riley}, \& {Allen}}]{Kouloumvakos2022}
{Kouloumvakos}, A., {Kwon}, R.~Y., {Rodr{\'\i}guez-Garc{\'\i}a}, L., {et~al.} 2022, \bibinfo{title}{{The first widespread solar energetic particle event of solar cycle 25 on 2020 November 29. Shock wave properties and the wide distribution of solar energetic particles},} \aap, 660, A84, \dodoi{10.1051/0004-6361/202142515}

\bibitem[{S. Krucker {et~al.}(2011)Krucker, Kontar, Christe, Glesener, \& Lin}]{Krucker2011}
Krucker, S., Kontar, E., Christe, S., Glesener, L., \& Lin, R. 2011, \bibinfo{title}{Electron acceleration associated with solar jets,} The Astrophysical Journal, 742, 82, \dodoi{10.1088/0004-637X/742/2/82}

\bibitem[{A. Kumar {et~al.}(2026)Kumar, Parashar, Kumar, Chakrabarty, Bapat, Sarkar, Shah, Adalja, Patel, Adhyaru, {et~al.}}]{Akumar2026}
Kumar, A., Parashar, S., Kumar, P., {et~al.} 2026, \bibinfo{title}{One year of ASPEX-SWIS operation: characteristic features, observations and science potential: A. Kumar et al.,} Journal of Astrophysics and Astronomy, 47, 46

\bibitem[{P. {Kumar} {et~al.}(2025){Kumar}, {Bapat}, {Shah}, {Adalja}, {Patel}, {Adhyaru}, {Shanmugam}, {Chakrabarty}, {Banerjee}, {Subramanian}, {Sarkar}, {Ladiya}, {Sebastian}, {Kumar}, {Kumar}, {Singh}, {Dadhania}, {Vadawale}, {Goyal}, {Tiwari}, {Sarda}, {Painkra}, {Sharma}, {Verma}, {Yogesh}, {Janardhan}, \& {Bhardwaj}}]{PKumar2025}
{Kumar}, P., {Bapat}, B., {Shah}, M.~S., {et~al.} 2025, \bibinfo{title}{{Aditya Solar Wind Particle Experiment (ASPEX) on Board Aditya{\textemdash}L1: The Solar Wind Ion Spectrometer (SWIS)},} \solphys, 300, 37, \dodoi{10.1007/s11207-025-02443-x}

\bibitem[{L.~D. Landau \& E.~M. Lifshitz(1959)Landau \& Lifshitz}]{LandauLifshitz1959}
Landau, L.~D., \& Lifshitz, E.~M. 1959, Course of Theoretical Physics, Vol.~6, Fluid Mechanics (Oxford: Pergamon Press)

\bibitem[{D. Lario {et~al.}(2006)Lario, Kallenrode, Decker, Roelof, Krimigis, Aran, \& Sanahuja}]{Lario2006}
Lario, D., Kallenrode, M.-B., Decker, R., {et~al.} 2006, \bibinfo{title}{Radial and longitudinal dependence of solar 4-13 MeV and 27-37 MeV proton peak intensities and fluences: Helios and IMP 8 observations,} The Astrophysical Journal, 653, 1531, \dodoi{10.1086/508982}

\bibitem[{R.~P. {Lepping} {et~al.}(1995){Lepping}, {Ac{\~{u}}na}, {Burlaga}, {Farrell}, {Slavin}, {Schatten}, {Mariani}, {Ness}, {Neubauer}, {Whang}, {Byrnes}, {Kennon}, {Panetta}, {Scheifele}, \& {Worley}}]{Lepping1995}
{Lepping}, R.~P., {Ac{\~{u}}na}, M.~H., {Burlaga}, L.~F., {et~al.} 1995, \bibinfo{title}{{The Wind Magnetic Field Investigation},} \ssr, 71, 207, \dodoi{10.1007/BF00751330}

\bibitem[{G. Li {et~al.}(2020)Li, Wu, Zhao, \& Yao}]{Li2020}
Li, G., Wu, X., Zhao, L., \& Yao, S. 2020, \bibinfo{title}{Observations of outward-propagating and mirroring of the same energetic electrons by wind,} The Astrophysical Journal Letters, 905, L1, \dodoi{10.1088/0004-637X/750/2/146}

\bibitem[{D.~J.~F. {Maia} \& M. {Pick}(2004){Maia} \& {Pick}}]{Maia2004}
{Maia}, D. J.~F., \& {Pick}, M. 2004, \bibinfo{title}{{Revisiting the Origin of Impulsive Electron Events: Coronal Magnetic Restructuring},} \apj, 609, 1082, \dodoi{10.1086/386319}

\bibitem[{O.~E. Malandraki {et~al.}(2002)Malandraki, Sarris, Lanzerotti, Trochoutsos, Tsiropoula, \& Pick}]{Malandraki2002}
Malandraki, O.~E., Sarris, E., Lanzerotti, L., {et~al.} 2002, \bibinfo{title}{Solar energetic particles inside a coronal mass ejection event observed with the ACE spacecraft,} Journal of atmospheric and solar-terrestrial physics, 64, 517, \dodoi{10.1016/S1364-6826(02)00008-1}

\bibitem[{W.~H. Matthaeus \& M.~L. Goldstein(1982)Matthaeus \& Goldstein}]{Matthaeus1982}
Matthaeus, W.~H., \& Goldstein, M.~L. 1982, \bibinfo{title}{Measurement of the rugged invariants of magnetohydrodynamic turbulence in the solar wind,} Journal of Geophysical Research: Space Physics, 87, 6011

\bibitem[{W.~H. Matthaeus {et~al.}(2003)Matthaeus, Qin, Bieber, \& Zank}]{Matthaeus2003}
Matthaeus, W.~H., Qin, G., Bieber, J.~W., \& Zank, G.~P. 2003, \bibinfo{title}{Nonlinear collisionless perpendicular diffusion of charged particles,} The Astrophysical Journal, 590, \dodoi{10.1086/376613}

\bibitem[{J. Mazur {et~al.}(2000)Mazur, Mason, Dwyer, Giacalone, Jokipii, \& Stone}]{Mazur2000}
Mazur, J., Mason, G., Dwyer, J., {et~al.} 2000, \bibinfo{title}{Interplanetary magnetic field line mixing deduced from impulsive solar flare particles,} The Astrophysical Journal Letters, 532, L79, \dodoi{10.1086/312561}

\bibitem[{T. {Nieves-Chinchilla} {et~al.}(2019){Nieves-Chinchilla}, {Jian}, {Balmaceda}, {Vourlidas}, {dos Santos}, \& {Szabo}}]{Chinchilla2019}
{Nieves-Chinchilla}, T., {Jian}, L.~K., {Balmaceda}, L., {et~al.} 2019, \bibinfo{title}{{Unraveling the Internal Magnetic Field Structure of the Earth-directed Interplanetary Coronal Mass Ejections During 1995 - 2015},} \solphys, 294, 89, \dodoi{10.1007/s11207-019-1477-8}

\bibitem[{K.~W. {Ogilvie} {et~al.}(1995){Ogilvie}, {Chornay}, {Fritzenreiter}, {Hunsaker}, {Keller}, {Lobell}, {Miller}, {Scudder}, {Sittler}, {Torbert}, \& et~al.}]{Ogilvie1995}
{Ogilvie}, K.~W., {Chornay}, D.~J., {Fritzenreiter}, R.~J., {et~al.} 1995, \bibinfo{title}{{SWE, A Comprehensive Plasma Instrument for the Wind Spacecraft},} \ssr, 71, 55, \dodoi{10.1007/BF00751326}

\bibitem[{M. {Paassilta} {et~al.}(2018){Paassilta}, {Papaioannou}, {Dresing}, {Vainio}, {Valtonen}, \& {Heber}}]{Paassilta2018}
{Paassilta}, M., {Papaioannou}, A., {Dresing}, N., {et~al.} 2018, \bibinfo{title}{{Catalogue of \{>\} 55 MeV Wide-longitude Solar Proton Events Observed by SOHO, ACE, and the STEREOs at \{{\ensuremath{\approx}}\} 1 AU During 2009 - 2016},} \solphys, 293, 70, \dodoi{10.1007/s11207-018-1284-7}

\bibitem[{S. {Pal} {et~al.}(2025){Pal}, {Mac Cormack}, {Kilpua}, {Yogesh}, {Jian}, \& {Nieves-Chinchilla}}]{Pal2025}
{Pal}, S., {Mac Cormack}, C., {Kilpua}, E. K.~J., {et~al.} 2025, \bibinfo{title}{{Magnetic interaction analysis of multiple interplanetary coronal mass ejections that led to a historic geomagnetic storm in May 2024},} \aap, 702, A150, \dodoi{10.1051/0004-6361/202555908}

\bibitem[{C. Palmroos {et~al.}(2025)Palmroos, Dresing, Gieseler, Guti{\'e}rrez, \& Vainio}]{Palmroos2025}
Palmroos, C., Dresing, N., Gieseler, J., Guti{\'e}rrez, C.~P., \& Vainio, R. 2025, \bibinfo{title}{A new method for determining the onset times of solar energetic particles and their uncertainties: Poisson-CUSUM bootstrap hybrid method,} Astronomy \& Astrophysics, 694, A221, \dodoi{10.1051/0004-6361/202451280}

\bibitem[{A. {Papaioannou} {et~al.}(2014){Papaioannou}, {Malandraki}, {Dresing}, {Heber}, {Klein}, {Vainio}, {Rodr{\'\i}guez-Gas{\'e}n}, {Klassen}, {Nindos}, {Heynderickx}, {Mewaldt}, {G{\'o}mez-Herrero}, {Vilmer}, {Kouloumvakos}, {Tziotziou}, \& {Tsiropoula}}]{Papaioannou2014}
{Papaioannou}, A., {Malandraki}, O.~E., {Dresing}, N., {et~al.} 2014, \bibinfo{title}{{SEPServer catalogues of solar energetic particle events at 1 AU based on STEREO recordings: 2007-2012},} \aap, 569, A96, \dodoi{10.1051/0004-6361/201323336}

\bibitem[{S. Parashar {et~al.}(2026)Parashar, Chakrabarty, Kumar, Kumar, Bapat, Sarkar, Janardhan, Bhardwaj, Vadawale, Shah, {et~al.}}]{Parashar2026}
Parashar, S., Chakrabarty, D., Kumar, P., {et~al.} 2026, \bibinfo{title}{Evidence for In Situ Particle Energization during the 2024 May Event Based on the ASPEX Instrument on Board Aditya-L1,} The Astrophysical Journal Letters, 996, L36, \dodoi{10.3847/2041-8213/ae2ac9}

\bibitem[{N.~P. Parate {et~al.}(2025)Parate, Shaji, Srikanth, Sankarasubramanian, Patnaik, Narendra, Adoni, Negi, Pattanaik, GV~P, {et~al.}}]{Parate2025}
Parate, N.~P., Shaji, N., Srikanth, M., {et~al.} 2025, \bibinfo{title}{Aditya-L1-An Observatory Class Mission: Spacecraft, Mission, and Operations,} Solar Physics, 300, 128

\bibitem[{G. Paschmann \& S.~J. Schwartz(2000)Paschmann \& Schwartz}]{PaschmannSchwartz2000}
Paschmann, G., \& Schwartz, S.~J. 2000, \bibinfo{title}{Five Lectures on the Plasma Physics of the Solar System,} in Analysis Methods for Multi-Spacecraft Data, ed. G.~Paschmann \& P.~W. Daly (Bern, Switzerland: ISSI)

\bibitem[{E.~R. Priest(2014)Priest}]{Priest2014}
Priest, E.~R. 2014, Magnetohydrodynamics of the Sun, 2nd edn. (Cambridge: Cambridge University Press), \dodoi{10.1017/CBO9781139020732}

\bibitem[{G. Qin {et~al.}(2004)Qin, Zhang, Dwyer, \& Rassoul}]{Qin2004}
Qin, G., Zhang, M., Dwyer, J.~R., \& Rassoul, H.~K. 2004, \bibinfo{title}{Interplanetary transport mechanisms of solar energetic particles,} The Astrophysical Journal, 609, 1076, \dodoi{10.1086/421101}

\bibitem[{D.~V. {Reames}(1999){Reames}}]{Reames1999}
{Reames}, D.~V. 1999, \bibinfo{title}{{Particle acceleration at the Sun and in the heliosphere},} \ssr, 90, 413, \dodoi{10.1023/A:1005105831781}

\bibitem[{D.~V. {Reames} {et~al.}(1996){Reames}, {Barbier}, \& {Ng}}]{Reames1996}
{Reames}, D.~V., {Barbier}, L.~M., \& {Ng}, C.~K. 1996, \bibinfo{title}{{The Spatial Distribution of Particles Accelerated by Coronal Mass Ejection--driven Shocks},} \apj, 466, 473, \dodoi{10.1086/177525}

\bibitem[{D.~V. {Reames} {et~al.}(1992){Reames}, {Richardson}, \& {Wenzel}}]{Reames1992}
{Reames}, D.~V., {Richardson}, I.~G., \& {Wenzel}, K.-P. 1992, \bibinfo{title}{{Energy Spectra of Ions from Impulsive Solar Flares},} \apj, 387, 715, \dodoi{10.1086/171120}

\bibitem[{I.~G. {Richardson} {et~al.}(2014){Richardson}, {von Rosenvinge}, {Cane}, {Christian}, {Cohen}, {Labrador}, {Leske}, {Mewaldt}, {Wiedenbeck}, \& {Stone}}]{Richardson2014}
{Richardson}, I.~G., {von Rosenvinge}, T.~T., {Cane}, H.~V., {et~al.} 2014, \bibinfo{title}{{> 25 MeV Proton Events Observed by the High Energy Telescopes on the STEREO A and B Spacecraft and/or at Earth During the First {\ensuremath{\sim}} Seven Years of the STEREO Mission},} \solphys, 289, 3059, \dodoi{10.1007/s11207-014-0524-8}

\bibitem[{L. {Rodr{\'\i}guez-Garc{\'\i}a} {et~al.}(2021){Rodr{\'\i}guez-Garc{\'\i}a}, {G{\'o}mez-Herrero}, {Zouganelis}, {Balmaceda}, {Nieves-Chinchilla}, {Dresing}, {Dumbovi{\'c}}, {Nitta}, {Carcaboso}, {dos Santos}, {Jian}, {Mays}, {Williams}, \& {Rodr{\'\i}guez-Pacheco}}]{RodriguezGarcia2021}
{Rodr{\'\i}guez-Garc{\'\i}a}, L., {G{\'o}mez-Herrero}, R., {Zouganelis}, I., {et~al.} 2021, \bibinfo{title}{{The unusual widespread solar energetic particle event on 2013 August 19. Solar origin and particle longitudinal distribution},} \aap, 653, A137, \dodoi{10.1051/0004-6361/202039960}

\bibitem[{L. {Rodr{\'\i}guez-Garc{\'\i}a} {et~al.}(2025){Rodr{\'\i}guez-Garc{\'\i}a}, {G{\'o}mez-Herrero}, {Dresing}, {Balmaceda}, {Palmerio}, {Kouloumvakos}, {Jebaraj}, {Espinosa Lara}, {Roco}, {Palmroos}, {Warmuth}, {Nicolaou}, {Mason}, {Guo}, {Laitinen}, {Cernuda}, {Nieves-Chinchilla}, {Fedeli}, {Lee}, {Cohen}, {Owen}, {Ho}, {Malandraki}, {Vainio}, \& {Rodr{\'\i}guez-Pacheco}}]{RodriguezGarcia2025}
{Rodr{\'\i}guez-Garc{\'\i}a}, L., {G{\'o}mez-Herrero}, R., {Dresing}, N., {et~al.} 2025, \bibinfo{title}{{Solar energetic particles injected inside and outside a magnetic cloud: The widespread solar energetic particle event on 2022 January 20},} \aap, 694, A64, \dodoi{10.1051/0004-6361/202452158}

\bibitem[{D. Ruffolo {et~al.}(2012)Ruffolo, Pianpanit, Matthaeus, \& Chuychai}]{Ruffolo2012}
Ruffolo, D., Pianpanit, T., Matthaeus, W., \& Chuychai, P. 2012, \bibinfo{title}{Random ballistic interpretation of nonlinear guiding center theory,} The Astrophysical Journal Letters, 747, L34, \dodoi{10.1088/2041-8205/747/2/L34}

\bibitem[{T. Sanderson {et~al.}(1985)Sanderson, Reinhard, Van~Nes, \& Wenzel}]{Sanderson1985}
Sanderson, T., Reinhard, R., Van~Nes, P., \& Wenzel, K.-P. 1985, \bibinfo{title}{Observations of three-dimensional anisotropies of 35-to 1000-keV protons associated with interplanetary shocks,} Journal of Geophysical Research: Space Physics, 90, 19, \dodoi{10.1029/JA090iA01p00019}

\bibitem[{A. Savitzky \& M.~J.~E. Golay(1964)Savitzky \& Golay}]{SavitzkyGolay1964}
Savitzky, A., \& Golay, M. J.~E. 1964, \bibinfo{title}{Smoothing and Differentiation of Data by Simplified Least Squares Procedures,} Analytical Chemistry, 36, 1627, \dodoi{10.1021/ac60214a047}

\bibitem[{R. Schlickeiser(2002)Schlickeiser}]{Schlickeiser2002}
Schlickeiser, R. 2002, Cosmic Ray Astrophysics, Astronomy and Astrophysics Library (Berlin, Heidelberg: Springer), \dodoi{10.1007/978-3-662-04814-6}

\bibitem[{S.~J. Schwartz(1998)Schwartz}]{Schwartz1998}
Schwartz, S.~J. 1998, \bibinfo{title}{Shock and discontinuity normals, Mach numbers, and related parameters,} ISSI Scientific Reports Series, 1, 249

\bibitem[{J. Sebastian {et~al.}(2025)Sebastian, Kumar, Chakrabarty, Kumar, Goyal, Vadawale, Shanmugam, Hasan, Sarkar, Dalal, Gupta, Parashar, Bapat, Shah, Adhyaru, Patel, Adalja, Tiwari, Sarda, Sharma, Ladiya, Kumar, Singh, Painkra, Verma, Banerjee, Subramanian, Dadhania, Janardhan, \& Bhardwaj}]{Jsebastian_et_al_2025}
Sebastian, J., Kumar, A., Chakrabarty, D., {et~al.} 2025, \bibinfo{title}{Comparison of Solar Wind flux and bulk parameters obtained from Aditya L1-ASPEX with Wind-3DP PESA-L and ACE-EPAM-LEMS120,} GSICS Quarterly, 18, \dodoi{10.25923/gmzc-9a28}

\bibitem[{J. Sebastian {et~al.}(2026)Sebastian, Dalal, Gupta, Goyal, Chakrabarty, Vadawale, Shanmugam, Tiwari, Patel, Sarkar, {et~al.}}]{Sebastian2026}
Sebastian, J., Dalal, B., Gupta, A., {et~al.} 2026, \bibinfo{title}{One Year of ASPEX-STEPS Operation: Characteristic Features, Observations and Science Potential,} Journal of Astrophysics and Astronomy, 47, 21, \dodoi{10.1007/s12036-026-10134-7}

\bibitem[{S. Seetha \& S. Megala(2017)Seetha \& Megala}]{Seetha2017}
Seetha, S., \& Megala, S. 2017, \bibinfo{title}{Aditya-L1 mission,} Current Science, 610

\bibitem[{A. Shalchi(2010)Shalchi}]{Shalchi2010}
Shalchi, A. 2010, \bibinfo{title}{A unified particle diffusion theory for cross-field scattering: subdiffusion, recovery of diffusion, and diffusion in three-dimensional turbulence,} The Astrophysical Journal Letters, 720, L127, \dodoi{10.1088/2041-8205/720/2/L127}

\bibitem[{G.~M. {Simnett} {et~al.}(2002){Simnett}, {Roelof}, \& {Haggerty}}]{Simnett2002}
{Simnett}, G.~M., {Roelof}, E.~C., \& {Haggerty}, D.~K. 2002, \bibinfo{title}{{The Acceleration and Release of Near-relativistic Electrons by Coronal Mass Ejections},} \apj, 579, 854, \dodoi{10.1086/342871}

\bibitem[{B.~U.~{\"O}. Sonnerup {et~al.}(2006)Sonnerup, Hasegawa, Teh, \& Hau}]{SonnerupEtAl2006}
Sonnerup, B. U.~{\"O}., Hasegawa, H., Teh, W.-L., \& Hau, L.-N. 2006, \bibinfo{title}{Grad-Shafranov reconstruction: An overview,} Journal of Geophysical Research: Space Physics, 111, \dodoi{10.1029/2006JA011717}

\bibitem[{B.~U.~O. Sonnerup \& M. Scheible(1998)Sonnerup \& Scheible}]{SonnerupScheible1998}
Sonnerup, B. U.~O., \& Scheible, M. 1998, \bibinfo{title}{Minimum and Maximum Variance Analysis,} in Analysis Methods for Multi-Spacecraft Data, ed. G.~Paschmann \& P.~W. Daly (Noordwijk, The Netherlands: European Space Agency), 185--220

\bibitem[{N.~S. Subashchandar {et~al.}(2025)Subashchandar, Zhao, Shalchi, Zank, Le~Roux, Li, Zhu, Silwal, \& Guzman}]{Subashchandar2025}
Subashchandar, N.~S., Zhao, L., Shalchi, A., {et~al.} 2025, \bibinfo{title}{Parallel and perpendicular diffusion of energetic particles in the near-Sun solar wind observed by Parker Solar Probe,} The Astrophysical Journal Letters, 991, L30, \dodoi{10.3847/2041-8213/ae063f}

\bibitem[{L.~C. Tan {et~al.}(2012)Tan, Malandraki, Reames, Ng, Wang, \& Dorrian}]{Tan2012}
Tan, L.~C., Malandraki, O.~E., Reames, D.~V., {et~al.} 2012, \bibinfo{title}{Use of incident and reflected solar particle beams to trace the topology of magnetic clouds,} The Astrophysical Journal, 750, 146, \dodoi{10.1088/0004-637X/750/2/146}

\bibitem[{G.~I. Taylor(1938)Taylor}]{Taylor1938}
Taylor, G.~I. 1938, \bibinfo{title}{The Spectrum of Turbulence,} Proceedings of the Royal Society of London. Series A, Mathematical and Physical Sciences, 164, 476, \dodoi{10.1098/rspa.1938.0032}

\bibitem[{L. {Trenchi} {et~al.}(2013{\natexlab{a}}){Trenchi}, {Bruno}, {D'Amicis}, {Marcucci}, \& {Telloni}}]{Trenchi2013a}
{Trenchi}, L., {Bruno}, R., {D'Amicis}, R., {Marcucci}, M.~F., \& {Telloni}, D. 2013{\natexlab{a}}, \bibinfo{title}{{Observations of IMF coherent structures and their relationship to SEP dropout events},} Annales Geophysicae, 31, 1333, \dodoi{10.5194/angeo-31-1333-2013}

\bibitem[{L. {Trenchi} {et~al.}(2013{\natexlab{b}}){Trenchi}, {Bruno}, {Telloni}, {D'amicis}, {Marcucci}, {Zurbuchen}, \& {Weberg}}]{Trenchi2013b}
{Trenchi}, L., {Bruno}, R., {Telloni}, D., {et~al.} 2013{\natexlab{b}}, \bibinfo{title}{{Solar Energetic Particle Modulations Associated with Coherent Magnetic Structures},} \apj, 770, 11, \dodoi{10.1088/0004-637X/770/1/11}

\bibitem[{R.~A. Treumann {et~al.}(2019)Treumann, Baumjohann, \& Narita}]{Treumann2019}
Treumann, R.~A., Baumjohann, W., \& Narita, Y. 2019, \bibinfo{title}{On the Applicability of Taylor's Hypothesis in Streaming Magnetohydrodynamic Turbulence,} Earth, Planets and Space, 71, 41, \dodoi{10.1186/s40623-019-1021-y}

\bibitem[{D. Tripathi {et~al.}(2022)Tripathi, Chakrabarty, Nandi, Prasad, Ramaprakash, Shaji, Sankarasubramanian, Thampi, \& Yadav}]{Tripathi2022}
Tripathi, D., Chakrabarty, D., Nandi, A., {et~al.} 2022, \bibinfo{title}{The Aditya-L1 mission of ISRO,} Proceedings of the International Astronomical Union, 18, 17

\bibitem[{D. Trotta {et~al.}(2024)Trotta, Larosa, Nicolaou, Horbury, Matteini, {et~al.}}]{Trotta2024}
Trotta, D., Larosa, A., Nicolaou, G., {et~al.} 2024, \bibinfo{title}{Properties of an Interplanetary Shock Observed at 0.07 and 0.7 au by Parker Solar Probe and Solar Orbiter,} The Astrophysical Journal, 962, 147, \dodoi{10.3847/1538-4357/ad187d}

\bibitem[{D. Trotta {et~al.}(2022)Trotta, Vuorinen, Hietala, Horbury, Dresing, Gieseler, Kouloumvakos, Price, Valentini, Kilpua, \& Vainio}]{Trotta2022}
Trotta, D., Vuorinen, L., Hietala, H., {et~al.} 2022, \bibinfo{title}{Single-spacecraft techniques for shock parameters estimation: A systematic approach,} Frontiers in Astronomy and Space Sciences, 9, 1005672, \dodoi{10.3389/fspas.2022.1005672}

\bibitem[{R. {Vainio} {et~al.}(2009){Vainio}, {Desorgher}, {Heynderickx}, {Storini}, {Fl{\"u}ckiger}, {Horne}, {Kovaltsov}, {Kudela}, {Laurenza}, {McKenna-Lawlor}, {Rothkaehl}, \& {Usoskin}}]{Vainio2009}
{Vainio}, R., {Desorgher}, L., {Heynderickx}, D., {et~al.} 2009, \bibinfo{title}{{Dynamics of the Earth's Particle Radiation Environment},} \ssr, 147, 187, \dodoi{10.1007/s11214-009-9496-7}

\bibitem[{A. Vourlidas(2014)Vourlidas}]{Vourlidas2014}
Vourlidas, A. 2014, \bibinfo{title}{The flux rope nature of coronal mass ejections,} Plasma Physics and Controlled Fusion, 56, 064001, \dodoi{10.1088/0741-3335/56/6/064001}

\bibitem[{W. {Wei} {et~al.}(2024){Wei}, {Lee}, {Dresing}, {Khoo}, {Jian}, {Luhmann}, {Cohen}, {Fraschetti}, {Zhuang}, {Huang}, {Owen}, {Nicolaou}, {Rodr{\'\i}guez-Garc{\'\i}a}, {Palmerio}, {Jebaraj}, {Lynch}, \& {Carcaboso}}]{Wei2024}
{Wei}, W., {Lee}, C.~O., {Dresing}, N., {et~al.} 2024, \bibinfo{title}{{Very Large and Long-lasting Anisotropies Caused by Sunward Streaming Energetic Ions: Solar Orbiter and STEREO A Observations},} \apjl, 973, L52, \dodoi{10.3847/2041-8213/ad78df}

\bibitem[{P.~D. Welch(1967)Welch}]{Welch1967}
Welch, P.~D. 1967, \bibinfo{title}{The Use of Fast Fourier Transform for the Estimation of Power Spectra: A Method Based on Time Averaging Over Short, Modified Periodograms,} IEEE Transactions on Audio and Electroacoustics, 15, 70, \dodoi{10.1109/TAU.1967.1161901}

\bibitem[{G. {Wibberenz} \& H.~V. {Cane}(2006){Wibberenz} \& {Cane}}]{Wibberenz2006}
{Wibberenz}, G., \& {Cane}, H.~V. 2006, \bibinfo{title}{{Multi-Spacecraft Observations of Solar Flare Particles in the Inner Heliosphere},} \apj, 650, 1199, \dodoi{10.1086/506598}

\bibitem[{N. Wijsen {et~al.}(2023)Wijsen, Lario, S{\'a}nchez-Cano, Jebaraj, Dresing, Richardson, Aran, Kouloumvakos, Ding, Niemela, {et~al.}}]{Wijsen2023}
Wijsen, N., Lario, D., S{\'a}nchez-Cano, B., {et~al.} 2023, \bibinfo{title}{The effect of the ambient solar wind medium on a CME-driven shock and the associated gradual solar energetic particle event,} The Astrophysical Journal, 950, 172, \dodoi{10.3847/1538-4357/acd1ed}

\bibitem[{R.~F. {Wimmer-Schweingruber} {et~al.}(2023){Wimmer-Schweingruber}, {Berger}, {Kollhoff}, {K{\"u}hl}, {Heber}, {Yang}, {Heidrich-Meisner}, {Klassen}, {Gomez-Herrero}, {Rodriguez-Pacheco}, {Ho}, {Mason}, {Janitzek}, {Kouloumvakos}, {Wang}, {Warmuth}, {Lario}, {Carcaboso}, {Owen}, {Bu{\v{c}}{\'\i}k}, {Pacheco}, {Malandraki}, {Allen}, {Rodriguez}, {Shukhobodskaia}, {Espinosa Lara}, {Cernuda}, {B{\"o}ttcher}, {Eldrum}, {Fleth}, \& {Xu}}]{Wimmer2023}
{Wimmer-Schweingruber}, R.~F., {Berger}, L., {Kollhoff}, A., {et~al.} 2023, \bibinfo{title}{{Unusually long path length for a nearly scatter-free solar particle event observed by Solar Orbiter at 0.43 au},} \aap, 678, A98, \dodoi{10.1051/0004-6361/202346319}

\bibitem[{H. {Xie} {et~al.}(2019){Xie}, {St. Cyr}, {M{\"a}kel{\"a}}, \& {Gopalswamy}}]{Xie2019}
{Xie}, H., {St. Cyr}, O.~C., {M{\"a}kel{\"a}}, P., \& {Gopalswamy}, N. 2019, \bibinfo{title}{{Statistical Study on Multispacecraft Widespread Solar Energetic Particle Events During Solar Cycle 24},} Journal of Geophysical Research (Space Physics), 124, 6384, \dodoi{10.1029/2019JA026832}

\bibitem[{V.~K. {Yadav} {et~al.}(2025){Yadav}, {Vijaya}, {Krishnam Prasad}, {Srikar}, {Mahajan}, {Mallikarjun}, {Zamani}, {Satyanarayana Kumari}, {Lohar}, {Kandpal}, {Narendra}, {Rai}, {Adoni}, {Veeresha}, {T.~S.}, {Kalpana}, {Srivastava}, \& {Vichare}}]{Yadav2025}
{Yadav}, V.~K., {Vijaya}, Y., {Krishnam Prasad}, B., {et~al.} 2025, \bibinfo{title}{{The Fluxgate Magnetometer (MAG) on Board Aditya-L1 Spacecraft},} \solphys, 300, 33, \dodoi{10.1007/s11207-025-02440-0}

\bibitem[{L.-L. Zhao {et~al.}(2017)Zhao, Adhikari, Zank, Hu, \& Feng}]{Zhao2017}
Zhao, L.-L., Adhikari, L., Zank, G., Hu, Q., \& Feng, X. 2017, \bibinfo{title}{Cosmic ray diffusion tensor throughout the heliosphere derived from a nearly incompressible magnetohydrodynamic turbulence model,} The Astrophysical Journal, 849, 88, \dodoi{10.3847/1538-4357/aa932a}

\bibitem[{L.-L. Zhao {et~al.}(2018)Zhao, Adhikari, Zank, Hu, \& Feng}]{Zhao2018}
Zhao, L.-L., Adhikari, L., Zank, G., Hu, Q., \& Feng, X. 2018, \bibinfo{title}{Influence of the solar cycle on turbulence properties and cosmic-ray diffusion,} The Astrophysical Journal, 856, 94, \dodoi{10.3847/1538-4357/aab362}

\bibitem[{L.-L. Zhao {et~al.}(2019)Zhao, Zank, Chen, Hu, Le~Roux, Du, \& Adhikari}]{Zhao2019}
Zhao, L.-L., Zank, G., Chen, Y., {et~al.} 2019, \bibinfo{title}{Particle acceleration at 5 au associated with turbulence and small-scale magnetic flux ropes,} The Astrophysical Journal, 872, 4, \dodoi{10.3847/1538-4357/aafcb2}

\bibitem[{L.-L. Zhao {et~al.}(2020)Zhao, Zank, Adhikari, Hu, Kasper, Bale, Korreck, Case, Stevens, Bonnell, {et~al.}}]{Zhao2020}
Zhao, L.-L., Zank, G., Adhikari, L., {et~al.} 2020, \bibinfo{title}{Identification of magnetic flux ropes from parker solar probe observations during the first encounter,} The Astrophysical Journal Supplement Series, 246, 26, \dodoi{10.3847/1538-4365/ab4ff1}

\bibitem[{L.-L. {Zhao} {et~al.}(2021{\natexlab{a}}){Zhao}, {Zank}, {Hu}, {Telloni}, {Chen}, {Adhikari}, {Nakanotani}, {Kasper}, {Huang}, {Bale}, {Korreck}, {Case}, {Stevens}, {Bonnell}, {Dudok de Wit}, {Goetz}, {Harvey}, {MacDowall}, {Malaspina}, {Pulupa}, {Larson}, {Livi}, {Whittlesey}, {Klein}, \& {Raouafi}}]{Zhao2021a}
{Zhao}, L.-L., {Zank}, G.~P., {Hu}, Q., {et~al.} 2021{\natexlab{a}}, \bibinfo{title}{{Detection of small magnetic flux ropes from the third and fourth Parker Solar Probe encounters},} \aap, 650, A12, \dodoi{10.1051/0004-6361/202039298}

\bibitem[{L.-L. {Zhao} {et~al.}(2021{\natexlab{b}}){Zhao}, {Zank}, {He}, {Telloni}, {Hu}, {Li}, {Nakanotani}, {Adhikari}, {Kilpua}, {Horbury}, {O'Brien}, {Evans}, \& {Angelini}}]{Zhao2021b}
{Zhao}, L.-L., {Zank}, G.~P., {He}, J.~S., {et~al.} 2021{\natexlab{b}}, \bibinfo{title}{{Turbulence and wave transmission at an ICME-driven shock observed by the Solar Orbiter and Wind},} \aap, 656, A3, \dodoi{10.1051/0004-6361/202140450}

\bibitem[{L.-L. Zhao {et~al.}(2027)Zhao, Zhu, Jiao, Jin, Subashchandar, Zank, Silwal, McComas, Desai, Raptis, {et~al.}}]{Zhao2027}
Zhao, L.-L., Zhu, X., Jiao, Y., {et~al.} 2027, \bibinfo{title}{Persistent Sub-MeV Ion Enhancement Downstream of an Interplanetary Shock: Multispacecraft Observations near L1,} The Astrophysical Journal Supplement Series, 288, 5, \dodoi{10.3847/1538-4365/aeb300}

\end{thebibliography}

\end{document}